\documentclass[sigconf]{acmart}

\AtBeginDocument{%
  }

\copyrightyear{2026}
\acmYear{2026}
\setcopyright{cc}
\setcctype{by}
\acmConference[UIST '26]{The 39th Annual ACM Symposium on User Interface Software and Technology}{November 02--05, 2026}{Detroit, MI, USA}
\acmBooktitle{The 39th Annual ACM Symposium on User Interface Software and Technology (UIST '26), November 02--05, 2026, Detroit, MI, USA}
\acmDOI{10.1145/3830398.3830483}
\acmISBN{979-8-4007-2856-3/2026/11}

\usepackage{multirow}
\usepackage{pifont}
\usepackage{enumitem}

\usepackage[normalem]{ulem}
\newcommand{\myul}[1]{%
  {%
    \renewcommand{\ULthickness}{0.25pt}%
    \setlength{\ULdepth}{2pt}%
    \textit{\uline{#1}}%
  }%
}

\definecolor{mybl}{RGB}{100,100,100}
\newcommand{\mycolor}[1]{\textcolor{mybl}{#1}}

\begin{document}

\title{PartInteractor: Intent-Driven Part-Aware 3D Authoring for Continuous Co-Creation in XR}

\author{Jianan Jiang}
\affiliation{
  \institution{The Pennsylvania State University}
  \city{State College}
  \state{Pennsylvania}
  \country{USA}}
\email{jnjiang@psu.edu}
\thanks{See our \href{https://horizonjohn.github.io/PartInteractor/}{\textbf{P{\scriptsize{ROJECT}} P{\scriptsize{AGE}}}} for source code and implementation details.}

\author{Bin Li}
\affiliation{
  \institution{The Pennsylvania State University}
  \city{State College}
  \state{Pennsylvania}
  \country{USA}}
\email{binli@psu.edu}

\renewcommand{\shortauthors}{Jiang et al.}


\begin{abstract}

  As Extended Reality (XR) evolves into an immersive computing medium, interactive 3D authoring becomes essential for creative and functional workflows. However, existing generative XR systems produce monolithic outputs lacking explicit semantic structure, limiting post-generation control. We introduce PartInteractor, a representation-to-interaction framework that investigates how semantic part hierarchies can be incorporated into generative XR authoring, and exposed as first-class, directly manipulable units, turning one-shot prompt-to-object generation into continuous component-level co-creation. PartInteractor supports speech, sketch, and image inputs, integrating an LLM interpreter with a retrieval-generation strategy to scaffold user intent prior to 3D generation. Instead of producing monolithic objects, our system generates semantically decomposed 3D assets with explicit part hierarchies, enabling rich component-level interaction over object structure and composition. Our evaluations suggest that part-aware representation increases post-generation control and reduces reliance on whole-object regeneration, while intent scaffolding mitigates ambiguity and improves intent-result alignment, together supporting more expressive and controllable human-AI co-creation workflows. These results highlight part-aware representation and intent scaffolding as promising design considerations for future generative XR authoring systems.

\end{abstract}

\ccsdesc[500]{Human-centered computing~Interactive systems and tools}
\ccsdesc[500]{Human-centered computing~Virtual reality}
\ccsdesc[500]{Human-centered computing~Mixed / augmented reality}
\ccsdesc[500]{Computing methodologies~Artificial intelligence}

\keywords{Extended Reality, Generative Authoring, Large Language Model, 3D Generation, Human-AI Co-Creation}
\begin{teaserfigure}
  \includegraphics[width=\textwidth]{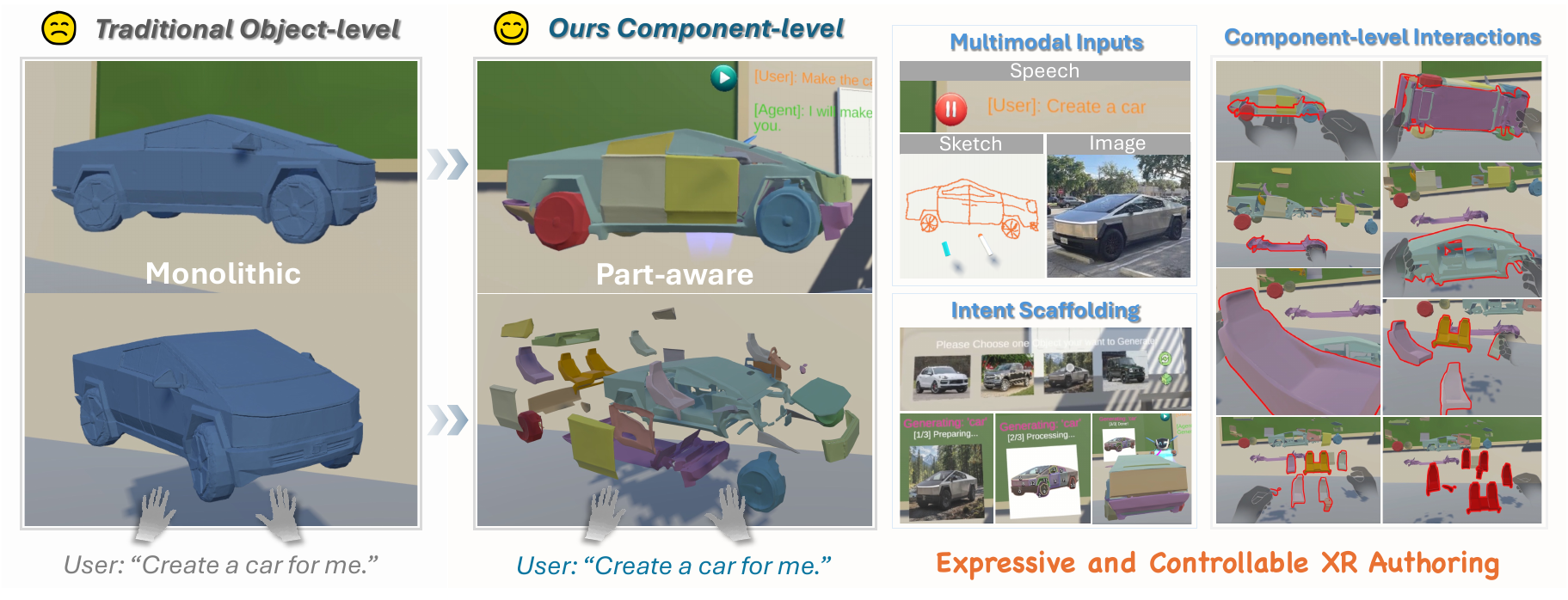}
  \vspace{-6mm}
  \caption{From object-level synthesis to intent-driven part-aware authoring in XR. \mycolor{(i) Existing XR generative systems synthesize monolithic objects that require full regeneration even for small adjustments. (ii) PartInteractor shifts this paradigm by integrating multimodal intent scaffolding with part-aware representation, producing semantically decomposed meshes with explicit part hierarchies that support fine-grained component-level interaction and enable continuous human-AI co-creation.}}
  \Description{Figure 1 illustrates the transition from traditional object-level XR generation to PartInteractor's component-level authoring workflow. A conventional system produces a monolithic car model that must be regenerated for changes, while PartInteractor produces a part-aware car with separate editable components. The right side shows multimodal input through speech, sketch, and image, intent scaffolding during generation, and XR interactions where users directly manipulate individual car parts.}
  \label{fig:teaser}
  \vspace{3mm}
\end{teaserfigure}


\maketitle

\section{Introduction}
\label{sec:introduction}

Over the past decades, Extended Reality (XR) has evolved from visualization systems~\cite{sutherland1968head, chung1989exploring} into powerful computing platforms~\cite{apple2023introducing, meta2025rayban} that support collaboration~\cite{ghamandi2024unlocking, numan2025cocreatar}, simulation~\cite{han2025spatio, caetano2025graspr}, and design~\cite{dogan2024augmented, lee20253d} within spatial environments. As XR transitions from delivering pre-authored experiences to enabling active production, the ability to create and modify 3D assets in immersive contexts becomes increasingly important~\cite{zhang2024vrcopilot, vachha2025dreamcrafter, lee2025imaginatear}. \textit{Beyond enabling in-environment authoring itself, XR systems must also support continuous authoring workflows, allowing users to iteratively create, adapt, and refine virtual objects without breaking embodied interaction.}

Recent advances in generative modeling~\cite{vaswani2017attention, brown2020language, ho2020denoising} have significantly lowered the barrier to 3D content creation. Multimodal approaches can synthesize assets from text, sketches, or images~\cite{xiang2025structured, tang2024lgm, xu2024grm, lee2025pasta}, enabling users to express high-level intent rather than perform manual modeling. Large language models (LLMs)~\cite{hurst2024gpt4o, grattafiori2024llama, songtang2025unrealllm, huang2025fireplace} further extend this capability by interpreting abstract instructions and reasoning about goal-directed actions. \textit{These developments open new opportunities for integrating generative intelligence into XR authoring workflows.}

However, most 3D generative pipelines~\cite{kerbl20233d, mildenhall2021nerf, tang2024lgm, xu2024grm} have been developed primarily for asset synthesis and reconstruction outside immersive, interaction-intensive authoring contexts. Existing generative XR authoring systems~\cite{zhang2024vrcopilot, vachha2025dreamcrafter, hu2025thing2reality, chen2025llmer} have made important progress in bringing generation into immersive environments, but typically treat generated content as object- or scene-level entities, as summarized in Table~\ref{tab:XR_comparison}. While this granularity is appropriate for many generation-centered XR workflows, immersive authoring often requires users to adjust individual components, inspect internal structures, or recombine parts across objects as ideas evolve. This creates a representation--interaction {\textbf{M\footnotesize{ISMATCH}}}: generated assets are presented at object or scene granularity, whereas post-generation XR authoring often requires finer-grained, component-level control. At the same time, although multimodal inputs such as speech, sketches, and images lower the barrier to expressing design intent, they remain inherently ambiguous~\cite{hu2025gesprompt, rosenberg2024drawtalking, shi2025caring}. Without mechanisms to interpret and scaffold intent prior to generation, early misunderstandings can propagate through the generative pipeline and make downstream refinement costly. \textit{These limitations suggest that generative XR authoring should move beyond object-level synthesis toward a workflow that scaffolds user intent before generation and preserves semantic part structure after generation as an interactable substrate.} Therefore, we identify three key challenges:

\begin{itemize}[leftmargin=8mm]
    \item[\textbf{C1}:] \textit{How can generative XR systems address the mismatch between generated object representations and interaction needs to enable more flexible and fine-grained user control?}

    \item[\textbf{C2}:] \textit{How can generative XR systems mitigate ambiguity in user input and system interpretation to support more effective and efficient intent expression?}
    
    \item[\textbf{C3}:] \textit{How can generative XR workflows support continuous and iterative authoring in immersive environments, rather than being limited to fragmented, generation-centric processes?}
\end{itemize}

\begin{table}[t]
\centering
\caption{Comparison of generative XR authoring systems on generated representation and interaction granularity.}
\Description{Table 1 compares five generative XR authoring systems on generated representation and interaction granularity. VRCopilot and Dreamcrafter generate object- or scene-level representations and support object- or scene-level interactions. Thing2Reality and LLMER generate object-level representations and support object-level interactions. In contrast, PartInteractor generates objects with explicit part hierarchies and supports both object-level and component-level interactions.}
\vspace{-2mm}
\label{tab:XR_comparison}
\setlength{\tabcolsep}{6pt}
\renewcommand{\arraystretch}{1.0}

\resizebox{\linewidth}{!}{
\begin{tabular}{lcc}
\toprule
\multirow{2}{*}{\textbf{System}} &
\textbf{Generated} &
\textbf{Interaction} \\
&
\textbf{Representation} &
\textbf{Granularity} \\
\midrule

VRCopilot~\cite{zhang2024vrcopilot} 
& Object / Scene
& Object-/Scene-level \\

Dreamcrafter~\cite{vachha2025dreamcrafter} 
& Object / Scene
& Object-/Scene-level \\

Thing2Reality~\cite{hu2025thing2reality} 
& Object
& Object-level \\

LLMER~\cite{chen2025llmer} 
& Object
& Object-level \\

\midrule
\textbf{PartInteractor (Ours)} 
& Object + Part Hierarchy
& Object-/Component-Level \\

\bottomrule
\end{tabular}
}

\vspace{-1mm}
\end{table}

To address these challenges, we present the \textit{PartInteractor} system that reframes XR authoring as a continuous human-AI co-creation workflow that scaffolds intent before generation and supports component-level interaction after generation. Unlike existing generative systems that produce monolithic objects from high-level prompts, PartInteractor generates assets with explicit semantic part hierarchies, allowing users to freely explore and refine objects and components. To support interpretable intent scaffolding, users can express design ideas through \textit{Speech} alone or in combination with \textit{in-situ Sketches} or \textit{Headset-Captured Images}. These multimodal inputs are processed by an LLM interpreter together with a retrieval-generation recommender to infer user intent. Instead of executing inferred intent immediately, PartInteractor adopts a human-in-the-loop workflow in which users can review, refine, and confirm 2D candidate suggestions before 3D generation. The confirmed visual candidate is then used to generate a semantically decomposed 3D asset that supports fine-grained component-level interaction in immersive environments. Through this design, \textit{PartInteractor shifts generative XR authoring from the single-shot prompt-to-object pipeline toward an intent-driven, part-aware workflow that supports continuous post-generation exploration and refinement}, in which intent formation, object generation, and interaction remain tightly coupled throughout the design process. We evaluate this approach through three user studies examining object manipulability, intent alignment, and system-level experience in XR authoring.

In summary, our contributions are as follows:

\begin{itemize}[leftmargin=2.8em]
    \item We introduce a \textbf{Representation-to-Interaction Framework} for generative XR authoring that preserves generated semantic part hierarchies and exposes them as first-class interaction units, enabling fine-grained component-level interaction beyond object-level regeneration.
    
    \item We introduce a human-in-the-loop \textbf{Intent-Scaffolding Workflow} that externalizes intent through user-confirmed visual candidates before 3D generation, reducing ambiguity and unnecessary generation attempts.
    
    \item We present \textbf{PartInteractor}, a multimodal XR authoring system that unifies pre-generation intent scaffolding, structured part-aware generation, and post-generation component-level interaction to support expressive and controllable human-AI co-creation.
    
    \item We evaluate our framework via two controlled studies and one system-level study, showing behavioral shifts toward component-level interaction, reduced whole-object regeneration, and improved perceived intent-result alignment.
\end{itemize}


\section{Related Work}
\subsection{Generative XR Authoring}
Recent advances in 3D generation and understanding have increasingly connected high-level user inputs with 3D representation, reconstruction, and generation~\cite{xue2023ulip, tang2024lgm, xu2024grm, xiang2025structured}. Within this broader progress, structured-shape methods have studied fine-grained part understanding~\cite{mo2019partnet, wang2026partnext}, hierarchical shape generation~\cite{mo2019structurenet, leng2024hypersdfusion}, procedural structure synthesis~\cite{jones2020shapeassembly}, and part-aware generation and decomposition~\cite{yang2025omnipart, yan2026x}. These works demonstrate the value of semantic structure for producing and organizing 3D assets. However, their focus remains largely on generation quality, structural coherence, or decomposition accuracy, rather than how generated structures can support downstream interaction workflows.

Although interactable structure has been explored in fabrication, mixed reality, and situated authoring~\cite{schulz2014design, faruqi2023style2fab, hu2023thingshare, arslan2025tinkerxr, iyer2025xr}, where users design, adapt, or share artifacts in relation to physical objects, materials, and environments, these systems typically focus on fabrication constraints, physical references, or CAD-like construction. In generative XR authoring, users must create objects from high-level intent while retaining the ability to interact with generated results within immersive environments. Recent XR systems have introduced generative models~\cite{haque2023instruct, xu2024instantmesh, tang2024lgm} and LLMs~\cite{hurst2024gpt4o, gpt2025image1} to lower this barrier, supporting in-situ asset generation or spatial content creation from language, images, or shared references~\cite{lee2025imaginatear, hu2025thing2reality, zhu2025agentar, numan2024spaceblender}, as well as prompt-based generation, direct manipulation, scene editing, and language-driven modeling~\cite{zhang2024vrcopilot, vachha2025dreamcrafter, de2024llmr, chen2025llmer}. These systems demonstrate the potential of generative XR authoring, but generated assets are still often treated as scene- or object-level entities, making fine-grained interaction difficult after generation.

In summary, prior work shows two complementary directions: generative 3D methods can produce diverse object entities, while generative XR systems enable immersive creation and interaction. Yet there remains a gap between generated representation and XR interaction granularity. Immersive design often requires users to explore and refine individual components as ideas evolve. \textit{Our work bridges this gap by introducing the part-aware representation that preserves and exposes semantic hierarchies after generation, enabling flexible and controllable component-level interaction.}

\vspace{-2mm}
\subsection{Human-AI Co-Creation}
Human-AI co-creation research has explored how AI collaborates with humans in creative and problem-solving tasks. Many systems position AI as an assistive tool augmenting workflows in domains such as coding~\cite{jonsson2022cracking, zhou2025instructpipe}, writing~\cite{yuan2022wordcraft, kim2023metaphorian}, and visual design~\cite{liu2022opal, jiang2024haigen}, while others emphasize iterative collaboration, where AI supports idea exploration, refinement, and feedback cycles~\cite{wang2024reelframer, shen2025ideationweb, suh2025storyensemble}. Similar paradigms also appear in education~\cite{prasad2025exploring, anderson2025exploring} and healthcare decision-making~\cite{lee2021human, zhang2024rethinking}, reflecting a shift toward AI as an interactive collaborator. As generative models enter XR authoring~\cite{zhang2024vrcopilot, de2024llmr, lee2025imaginatear, chen2025llmer, zhu2026generative}, interaction paradigms are moving toward intent-driven creation. However, interpreting user intent remains challenging due to ambiguous multimodal inputs. Several systems introduce intermediate representations or additional modalities to stabilize this process, including structured language-to-command representations~\cite{chen2025llmer}, gesture-, sketch-, and speech-based multimodal interaction~\cite{hu2025gesprompt, rosenberg2024drawtalking, duan2026justshape}, context-aware instruction authoring~\cite{shi2025caring}, embodied agent gestures for spatial communication~\cite{liu2026agenthands}, and progressive suggestions for iterative scene development~\cite{hou2025echoladder, aghel2024people}.

While these approaches improve intent interpretation and execution, generative XR authoring still requires outputs that support exploration and revision of incomplete ideas. \textit{We therefore introduce a human-in-the-loop intent-scaffolding pipeline that presents candidate visual specifications before 3D generation and supports multimodal inputs---including speech, in-situ sketches, and headset-captured images---allowing intent to be progressively shaped.}

\begin{figure}[t]
  \includegraphics[width=\linewidth]{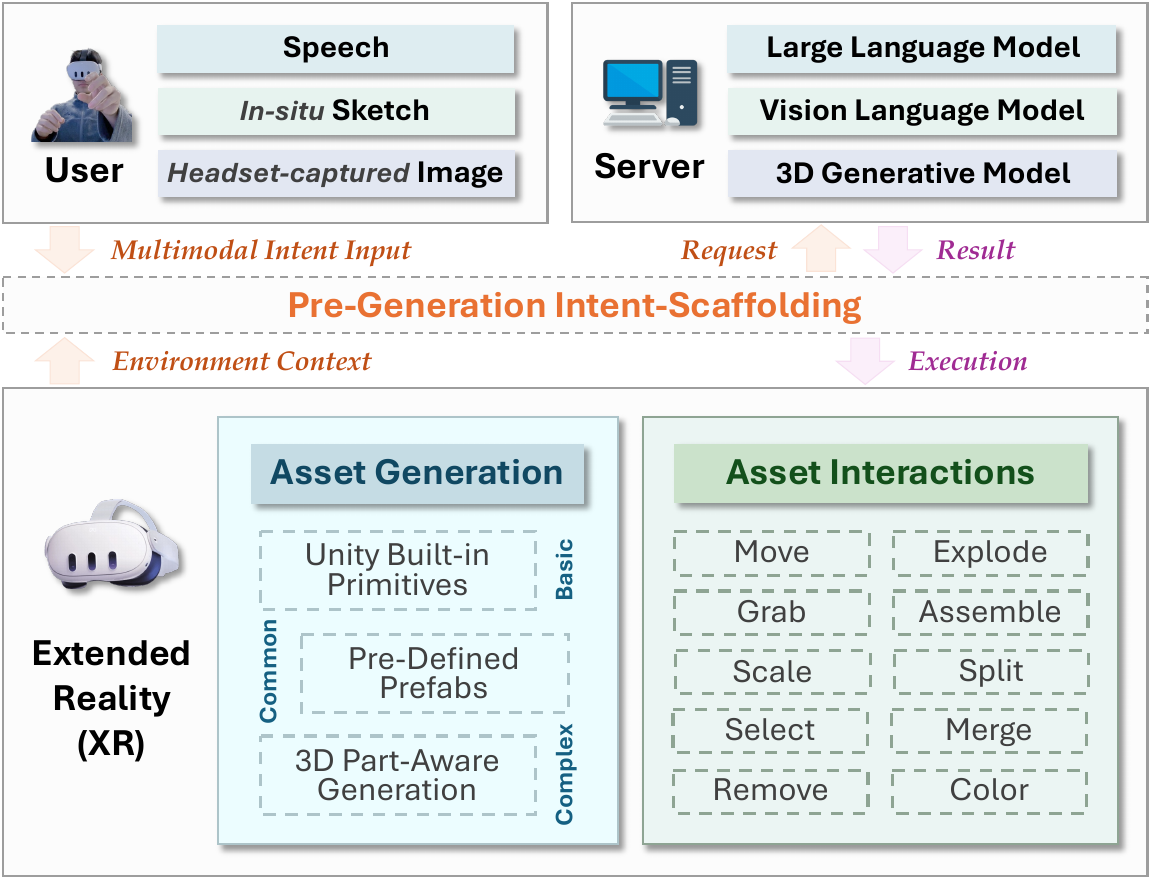}
  \vspace{-5mm}
  \caption{System overview. PartInteractor forms a continuous generative XR authoring workflow \mycolor{in which multimodal user intent and environmental context are used to scaffold intent before generation, followed by part-aware 3D asset generation and post-generation component-level interaction.}}
  \Description{Figure 2 shows the PartInteractor system workflow. Users provide speech, in-situ sketches, and headset-captured images as multimodal inputs. A server uses language, vision-language, and 3D generative models to scaffold user intent and support part-aware 3D asset generation. In the XR environment, the system supports asset generation from primitives, prefabs, or part-aware 3D generation, followed by component-level interactions such as moving, grabbing, scaling, selecting, removing, exploding, assembling, splitting, merging, and coloring parts.}
  \label{fig:system_architecture}
  \vspace{-2mm}
\end{figure}

\begin{figure*}[t]
  \includegraphics[width=\linewidth]{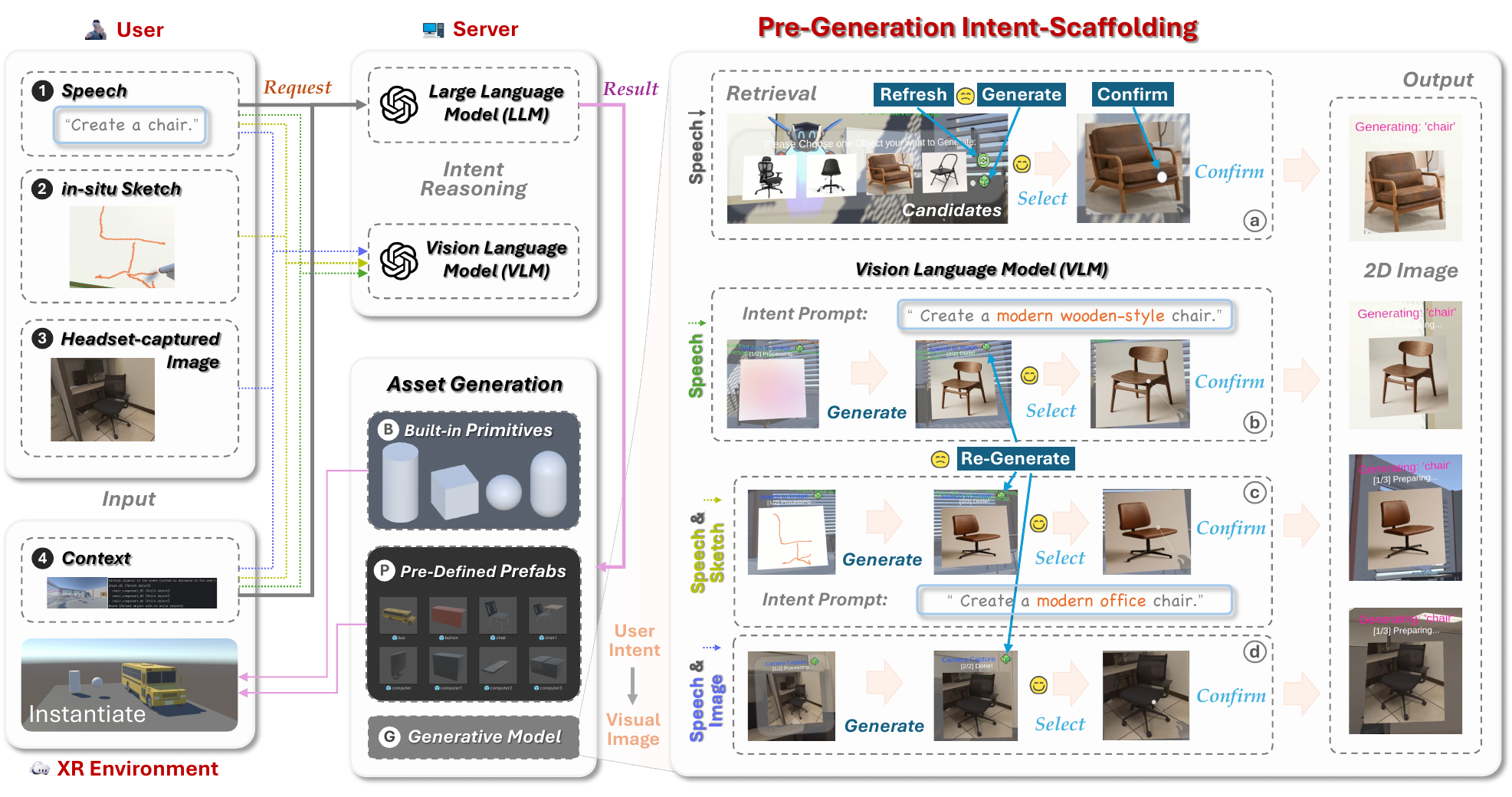}
  \vspace{-6mm}
  \caption{Pre-Generation Intent Scaffolding. \mycolor{User intent is scaffolded from multimodal inputs, including \textit{Speech}, \textit{in-situ Sketch}, \textit{Headset-captured Image}, and \textit{Environmental Context}, and presented as visual candidates. The system supports three pathways for expressing intent: (1) directly instantiating existing assets such as Primitives \textcolor{gray}{\textcircled{\footnotesize B}} and Prefabs \textcolor{gray}{\textcircled{\footnotesize P}}, (2) retrieving relevant objects \textcolor{gray}{\textcircled{a}} from local resources, and (3) generating candidates through a VLM \textcolor{gray}{\textcircled{b} \hspace{-1.5mm} \textcircled{c} \hspace{-1.5mm} \textcircled{d}} when suitable retrieved results are unavailable or unsatisfactory. This process allows users to explore and refine design intent before committing to 3D generation.}}
  \Description{Figure 3 illustrates PartInteractor's pre-generation intent scaffolding process. Users provide speech, in-situ sketches, headset-captured images, and XR environmental context as input. A server uses language and vision-language models to interpret the intent and connect it to asset generation options, including built-in primitives, predefined prefabs, and generative models. The right side shows four scaffolding pathways where users retrieve, generate, regenerate, select, and confirm 2D visual candidates before committing to 3D asset generation.}
  \label{fig:intent}
  \vspace{-1mm}
\end{figure*}

\section{PartInteractor System}
To address the challenges identified in Section~\ref{sec:introduction}, we derive a set of \textit{Design Goals} ($\S$~\ref{subsection:DG}) from the limitations of existing generative XR authoring systems. Guided by these goals, we design PartInteractor as a multimodal XR authoring system that reconceptualizes generative XR authoring as a human-in-the-loop workflow rather than a one-shot prompt-to-object operation. Specifically, PartInteractor tightly couples (i) \textit{Pre-Generation Intent Scaffolding}~($\S$~\ref{subsubsection:intent}), (ii) \textit{Structured Part-Aware Generation}~($\S$~\ref{subsubsection:generation}), and (iii) \textit{Post-Generation Component-Level Interaction}~($\S$~\ref{subsubsection:interaction}). This pipeline allows users to progressively externalize and refine their intent, produce structured part-aware assets, and flexibly interact with generated objects and their components within and across object boundaries.

\subsection{Design Goals}
\label{subsection:DG}

Guided by the challenges identified earlier, we formulated the following design goals for generative XR authoring systems:

\begin{itemize}
    \item[\textbf{DG1}:]
    \textit{Generated objects should maintain explicit semantic part hierarchies so that individual components remain available for direct interaction, rather than being collapsed into a single undifferentiated object.}

    \item[\textbf{DG2}:]
    \textit{The system should help users explore and clarify multimodal input before 3D generation, enabling them to shape system interpretation through a human-in-the-loop process rather than committing immediately to a generated result.}

    \item[\textbf{DG3}:]
    \textit{Generative XR authoring should support a continuous workflow in which users remain involved across intent formation, object generation, and subsequent interaction, enabling sustained component-level interaction and iterative co-creation.}
\end{itemize}

\subsection{Overview and Interaction Workflow}
We illustrate the architecture and interaction workflow of our PartInteractor system in Figure~\ref{fig:system_architecture}. The system forms a continuous XR authoring workflow that connects multimodal user input, generative model inference, and post-generation interaction within an immersive workspace. Users express design intent primarily through \textit{speech}, optionally complemented by \textit{in-situ sketches} or \textit{headset-captured images}. These multimodal inputs are interpreted by a backend pipeline that integrates a Large Language Model (LLM)~\cite{hurst2024gpt4o} and a Vision Language Model (VLM)~\cite{gpt2025image1}. Before committing to expensive 3D synthesis, the system presents 2D candidate visual suggestions that allow users to confirm or adjust system interpretations. Once a candidate is selected, a structured 3D asset is generated and instantiated within the XR workspace, enabling subsequent flexible and fine-grained interactions. This staged workflow allows users to shape intent before generation and maintain component-level control over generated content afterward, supporting expressive and controllable human-AI co-creation in XR.

\begin{figure*}[t]
  \includegraphics[width=\linewidth]{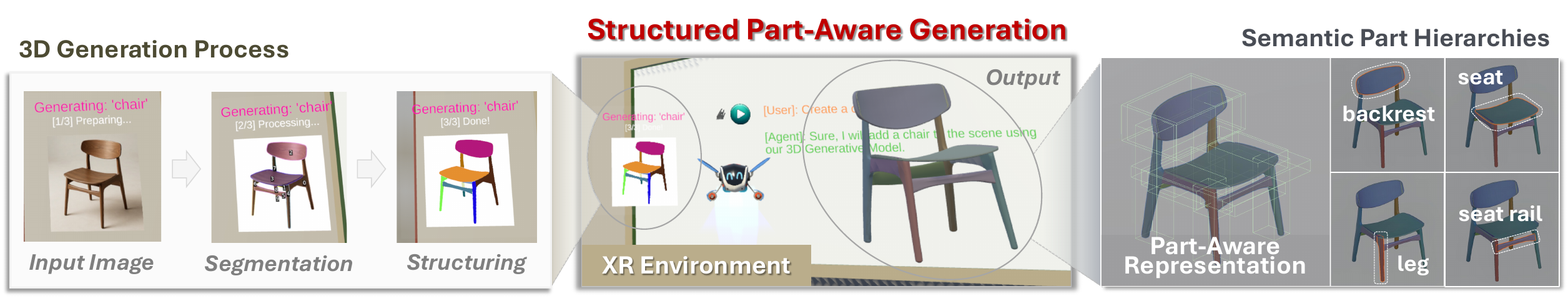}
  \vspace{-7mm}
  \caption{Structured Part-Aware Generation. \mycolor{Using the selected 2D image produced during the \textit{Pre-Generation Intent Scaffolding} stage, the system generates a structured 3D object through a multi-stage pipeline consisting of semantic segmentation and structural organization. The generated object is instantiated in the XR environment with preserved semantic part hierarchies (e.g., backrest, seat, leg, and seat rail), enabling subsequent flexible and fine-grained interactions.}}
  \Description{Figure 4 shows PartInteractor's structured part-aware generation process. A selected 2D chair image is transformed into a structured 3D object through segmentation and structural organization, then instantiated in the XR environment. The generated chair preserves a semantic part hierarchy with components such as the backrest, seat, legs, and seat rail, enabling later fine-grained component-level interaction.}
  \label{fig:generation}
  \vspace{-2mm}
\end{figure*}

\subsubsection{\textbf{Pre-Generation Intent Scaffolding}}
\label{subsubsection:intent}
Figure~\ref{fig:intent} illustrates our intent-scaffolding pipeline, which externalizes user intent before committing to 3D generation. Directly triggering 3D generation from an initial instruction~\cite{chen2025llmer} can be inefficient because unsatisfactory results often require repeated regeneration, while 3D synthesis remains computationally expensive and difficult to correct after generation. To address this limitation, our system introduces an intermediate candidate stage that generates lightweight visual suggestions before full 3D synthesis, allowing users to validate and adjust system interpretations of their intent at lower cost.

\textbf{Speech-Centered Multimodal Input.}
Users initiate object creation through spoken instructions as the primary modality. To further clarify design intent, users may optionally provide complementary visual cues, such as in-situ sketches or headset-captured images. These additional modalities provide structural or appearance references that help ground ambiguous descriptions.

\textbf{Intent Interpretation and Asset Routing.}
User instructions are interpreted by an LLM-based reasoning pipeline. Based on the interpreted intent, the system dynamically selects an appropriate asset creation strategy. For simple objects, built-in primitives (e.g., cubes or spheres) are instantiated directly. If relevant assets exist within the local resource library, the system loads matching prefabs. When no suitable asset is available, the system activates the retrieval-generation pipeline to provide candidate representations.

\textbf{Retrieval-Generation Strategy.}
Within this pipeline, PartInteractor employs a retrieval-generation module that presents candidate visual representations before invoking full 3D generation. Retrieval serves as a fast first-pass exploration strategy, while generation is used when retrieved results are unavailable or unsatisfactory. The system retrieves relevant objects from a local asset library and displays four candidate images in the XR interface, allowing users to review and select among different visual interpretations of their intent. If none of the retrieved candidates match the user’s intent, users can either refresh the results or trigger the generation-based method:
(i)~\myul{Speech-driven Generation}: synthesizing candidates directly from speech input;
(ii)~\myul{Sketch-guided Generation}: using in-situ sketches drawn on a whiteboard to provide structural cues;
(iii)~\myul{Image-guided Generation}: leveraging headset-captured images as visual references.

\textbf{Candidate Selection.}
Each generation request produces one candidate image representing a possible interpretation of the user’s intent. Users can iteratively explore and refine visual candidates until a satisfactory result is obtained. The selected image then serves as a 2D intent specification for the subsequent 3D generation stage.


\subsubsection{\textbf{Structured Part-Aware Generation}}
\label{subsubsection:generation}
Figure~\ref{fig:generation} presents our structured part-aware generation pipeline. After a visual candidate is selected during the \textit{Pre-Generation Intent Scaffolding} stage, PartInteractor converts the corresponding 2D representation into a structured 3D asset. While many existing generative authoring systems~\cite{vachha2025dreamcrafter, lee2025imaginatear} rely on 3D generation approaches that produce monolithic representations~\cite{haque2023instruct, xu2024instantmesh}, PartInteractor builds on OmniPart~\cite{yang2025omnipart} to obtain objects with explicit semantic part hierarchies, which are then preserved and exposed as editable units for subsequent component-level interactions.

Specifically, given a selected 2D representation, the model infers a semantic decomposition of the object and generates geometry for individual components while preserving their spatial relationships. The resulting asset maintains a hierarchy of functional parts (e.g., backrest, seat, legs, and seat rail), where components remain as separate entities rather than being merged into a single mesh. This structured representation retains both geometric coherence and semantic organization of the object.

Once generated, the structured object is instantiated within the XR environment. Because semantic part hierarchies are preserved, each component can be individually selected and interacted with during the subsequent \textit{Interaction} stage. By aligning the structure of representation with the granularity of interaction, this design enables flexible and fine-grained interactions and supports continuous human-AI co-creation in generative XR authoring workflows.


\begin{figure*}[t]
  \includegraphics[width=0.96\linewidth]{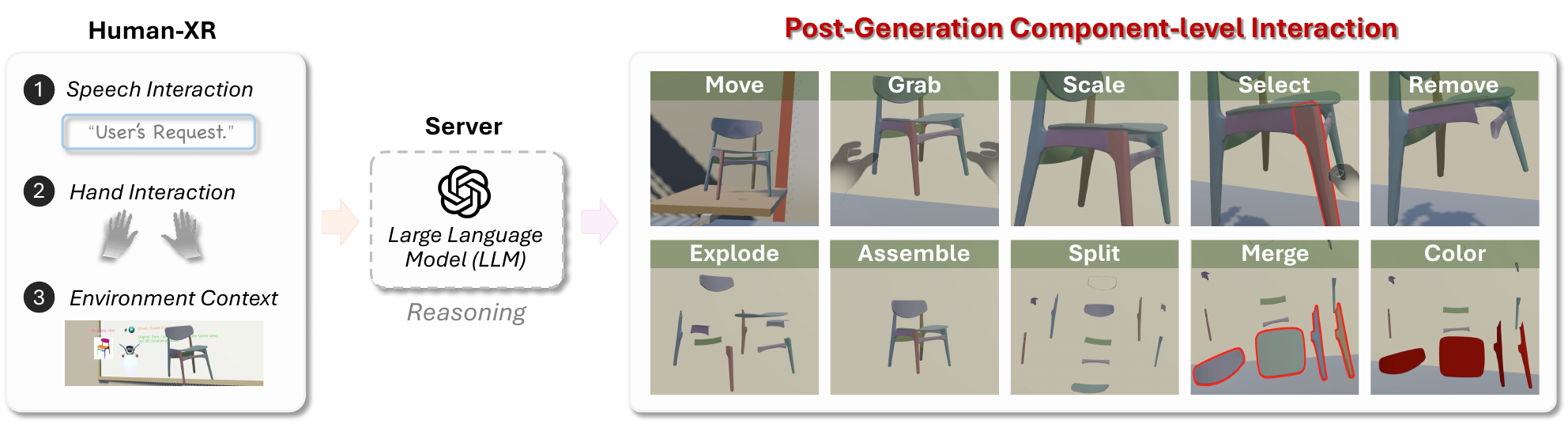}
  \vspace{-3mm}
  \caption{Post-Generation Component-Level Interaction. \mycolor{After a structured object is generated, users interact with individual object components through user instructions interpreted by an LLM with XR scene context. The preserved semantic part hierarchy enables fine-grained operations, including spatial manipulation (Move, Grab, Scale), component editing (Select, Remove), structural inspection (Explode, Assemble), structural modification (Split, Merge), and appearance customization (Color), allowing continuous human-AI co-creation without whole-object regeneration.}}
  \Description{Figure 5 shows PartInteractor's post-generation component-level interaction workflow. Users provide speech, hand input, and XR scene context, which are interpreted by a large language model. The system then supports flexible fine-grained interactions on individual chair components, including move, grab, scale, select, remove, explode, assemble, split, merge, and color.}
  \label{fig:interaction}
  \vspace{-2mm}
\end{figure*}

\subsubsection{\textbf{Post-Generation Component-Level Interaction}}
\label{subsubsection:interaction}
After a structured object is generated and instantiated in the XR environment, users can directly interact with individual components. As illustrated in Figure~\ref{fig:interaction}, the preserved semantic part hierarchy enables individual components to function as editable interaction units rather than fixed elements of a single mesh.

\textbf{Context-Aware Reasoning.}
PartInteractor interprets post-generation user actions through an LLM-based reasoning module grounded in user input and XR scene context. The system considers the user’s current instruction together with structured environmental information, including objects present in the scene, their semantic part hierarchies, spatial properties such as position and scale, and the currently referenced object or component. Dialogue history from previous interactions is also incorporated to maintain continuity across multi-step authoring sessions. Based on this contextual information, the LLM infers the intended interaction and produces a structured action command specifying the target object/component and the operation to be executed in the XR environment. This enables the system to resolve underspecified user requests, such as referring expressions and follow-up modifications, and to support flexible and fine-grained interactions without requiring explicit parameter specification.

\textbf{Flexible and Fine-Grained Interactions.}
Our system supports the following object-/component-level interactions within/across objects:
(i)~\myul{Spatial Manipulation:} Each object or component can be independently manipulated using spatial operations such as \textit{Move}, \textit{Grab}, and \textit{Scale}, allowing users to reposition or resize generated content at different granularity levels while preserving the overall object structure when needed.
(ii)~\myul{Component Editing:} Users can perform editing operations such as \textit{Select} and \textit{Remove} to isolate specific components or delete them from the object structure without affecting the remaining parts.
(iii)~\myul{Structural Inspection:} To facilitate structural understanding, the \textit{Explode} operation separates object components to expose internal structure, allowing users to inspect individual parts in isolation. The complementary \textit{Assemble} operation restores the object configuration after inspection.
(iv)~\myul{Structural Modification:} Beyond inspection, users can modify hierarchical relationships using \textit{Split} and \textit{Merge}. These operations allow components to be detached from their original parent structures and reattached to other components within or across generated objects, enabling flexible restructuring beyond whole-object editing.
(v)~\myul{Appearance Customization:} Users can also customize visual properties of generated objects or their individual components through appearance-level operations such as \textit{Color}, without modifying the underlying object hierarchy or geometry.

These interaction capabilities transform generated objects from static outputs into editable design artifacts. By combining part-aware representation with component-level interaction, PartInteractor enables users to iteratively manipulate, inspect, restructure, and customize generated content, supporting a continuous human-AI co-creation workflow in generative XR authoring.

\subsection{Implementation Details}

Our system runs on the Meta Quest 3~\footnote{\url{https://www.meta.com/quest/quest-3/}} and is developed in Unity~\footnote{\url{https://unity.com/products/unity-engine}} 2022.3.58f1 using the Meta All-in-One Interaction SDK and additional XR packages. The XR client supports in-situ speech input, sketch input, headset-captured images, hand-based interaction, and runtime manipulation of generated objects and components. The headset communicates with a local server through a socket-based protocol. The local server, equipped with an Intel i9-13900K CPU and an NVIDIA RTX 4070 GPU, acts as the middleware between the XR client and the cloud-based AI services. It receives user inputs and scene context from Unity, formats model requests, manages intermediate generation results, and returns structured commands or generated assets to the XR environment.

The cloud backend integrates several AI services, including GPT-4o~\cite{hurst2024gpt4o}, GPT-Image-1~\cite{gpt2025image1}, and OmniPart~\cite{yang2025omnipart}, which are accessed through external APIs to support the generative authoring pipeline. GPT-4o is used for command reasoning and action planning, including interpreting user requests, resolving referenced objects or components, and producing structured action commands for the XR client. For pre-generation intent scaffolding, the XR client first queries a self-collected local candidate library containing 10 object categories, each with 12 fine-grained 2D candidate images. The library is indexed by category tags extracted from user intent and can be extended by adding new tagged candidate images. When retrieved candidates are insufficient or when users provide sketch or image references, GPT-Image-1 generates additional 2D candidate images that serve as visual specifications for subsequent 3D generation. After the user confirms a candidate, OmniPart generates a part-aware 3D asset with explicit part structures. We use its default segmentation setting with a minimum segment size of 2000 pixels and no manually specified merge groups. The resulting segmentation map serves as guidance for part-aware 3D generation and initial part organization. The generated asset is imported into Unity as a GLB file using glTFast~\footnote{https://github.com/atteneder/glTFast}, while preserving the object hierarchy and part-level transforms so that individual components can be exposed as selectable and manipulable units in the XR workspace.

\begin{figure*}[t]
  \includegraphics[width=0.86\linewidth]{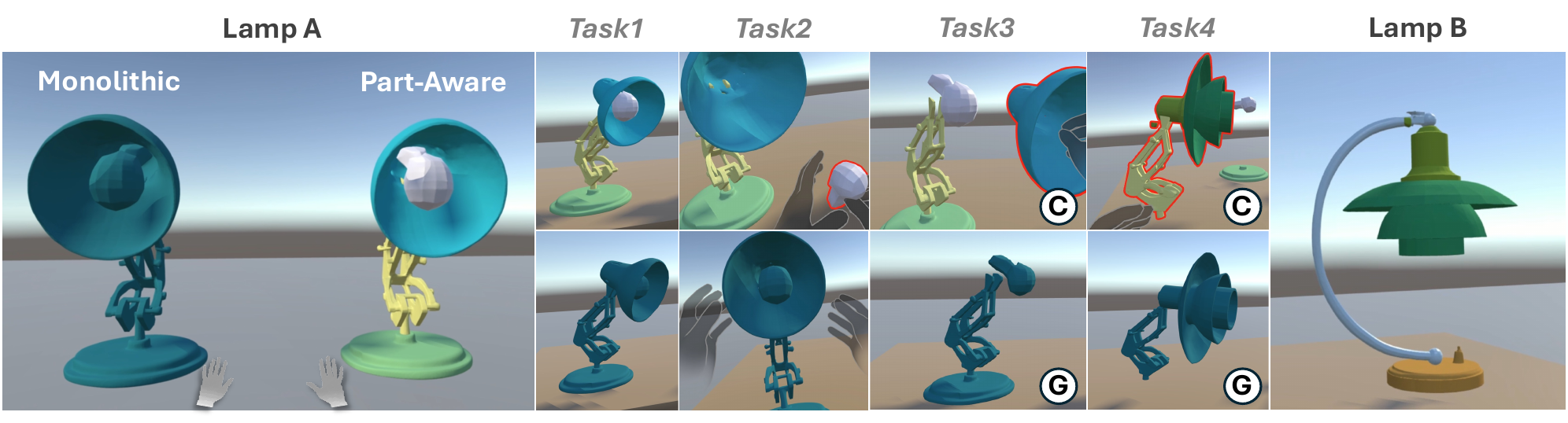}
  \vspace{-2mm}
  \caption{User Study on Object Manipulability. \mycolor{Tasks highlight two interaction strategies: \textcircled{\footnotesize C} represents component-level interaction enabled by part-aware representations, whereas \textcircled{\footnotesize G} indicates regeneration-based interaction.}}
  \Description{Figure 6 presents the object manipulability study using two desk lamp designs. Lamp A is shown as both a monolithic object and a part-aware object with accessible components. Four task examples illustrate sequential XR editing phases: placing the lamp, inspecting the lamp head and bulb, removing the lampshade while preserving other parts, and recombining Lamp A's arm with Lamp B's lampshade.}
  \label{fig:study1}
  \vspace{-1mm}
\end{figure*}

\section{Evaluation}
\subsection{Overview}

We conducted two controlled ablation studies and one complementary open-ended system-level study to evaluate how PartInteractor supports generative XR authoring workflows. Guided by our design challenges, the evaluation focuses on: \textit{Object Manipulability} (C1), \textit{Intent Alignment} (C2), and \textit{System-Level Co-Creation} (C3).

\subsubsection{\textbf{Hypotheses}}
The three studies examine complementary aspects of our system design: Study 1 and Study 2 isolate the effects of part-aware representation and intent scaffolding, while Study 3 examines the integrated workflow of PartInteractor in an open-ended authoring setting, focusing on how users appropriate intent scaffolding from multimodal input and component-level interaction during co-creation. Specifically, we test the following hypotheses:
\begin{itemize}
    \item[\textbf{H1}:] \textit{Part-aware representation improves users’ perceived control by enabling fine-grained component-level interaction, leading to reduced reliance on regeneration and a shift from object-level manipulation to direct structural editing.}
    
    \item[\textbf{H2}:] \textit{Intent scaffolding improves perceived alignment between user intent and generated results by reducing ambiguity prior to 3D generation, resulting in fewer regeneration attempts and faster task completion.}
    
    \item[\textbf{H3}:] \textit{Part-aware XR authoring with intent scaffolding will support sustained within-object interaction and positive user perceptions of engagement and expressiveness.}
\end{itemize}

\subsubsection{\textbf{Procedure}}

We employed a within-subject design in which all participants completed three studies in a single 60-70 minute session. The session began with a standardized training phase introducing the system’s interaction techniques. For studies involving multiple system conditions, we counterbalanced the condition order using a Latin-square scheme to mitigate order effects. After each study, participants completed a post-task questionnaire (7-point Likert scale) to assess their experience. For Studies 1--2, questionnaire responses were analyzed using paired-sample tests with mean differences ($MD$), 95\% confidence intervals ($CI_{95\%}$), Cohen's ($d_z$), and Holm-corrected ($p_{\mathrm{Holm}}$)-values. Q5--Q8 from the exploratory Study 3 are reported descriptively. We inserted short breaks of roughly three minutes between studies to mitigate fatigue. After completing all three studies, participants took part in a brief semi-structured interview to reflect on their overall experience.

\subsubsection{\textbf{Participants}}

We recruited 15 participants (5 female, 10 male; age range 20-30, mean = 24.7) from a university campus. All participants had normal or corrected-to-normal vision. Participants had diverse academic backgrounds spanning environmental sciences, mathematics, and engineering and computing disciplines. 9 participants reported prior XR experience, and 8 reported experience with 3D modeling tools. This study was approved by the IRB, and participants received a \$15 gift card as compensation.


\subsection{Study 1: Object Manipulability}

This study investigates \textbf{H1} by examining how different object representations shape post-generation interaction workflows.

\subsubsection{\textbf{Experimental Conditions}}

We compare two representational paradigms that support structural modification through different interaction workflows:
(i) \myul{Monolithic Representation} (Baseline): The semantic hierarchy was hidden, and all components were instantiated as a single mesh, such that structural changes were performed through whole-object regeneration.
(ii) \myul{Part-Aware Representation} (PartInteractor): The semantic hierarchy was accessible, enabling component-level interaction for structural modification. Whole-object regeneration and interaction were still available.

The baseline condition reflects existing generative XR pipelines in which generated assets are instantiated as monolithic objects without explicit semantic part hierarchies. To ensure that observed differences arise from the representational paradigm rather than content variability, both conditions used the same prefab resource pool, shared identical geometry, and system configurations.

\subsubsection{\textbf{Task Design}}

As shown in Figure~\ref{fig:study1}, participants completed a four-phase task involving iterative modification of a virtual desk lamp in XR. The task was designed to reflect common post-generation interaction needs in XR authoring workflows, ranging from object-level placement to target structural modification:

\begin{itemize}[leftmargin=12mm]
    \item[\textbf{Task 1}:] \myul{Object Placement}. Participants moved \textit{Lamp A} onto a designated region of the desk.
    \item[\textbf{Task 2}:] \myul{Structural Inspection}. Participants inspected \textit{Lamp A}’s head and examined the bulb detail.
    \item[\textbf{Task 3}:] \myul{Component Removal}. Participants selected and removed \textit{Lamp A}’s lampshade while preserving remaining parts.
    \item[\textbf{Task 4}:] \myul{Cross-Object Recombination}. Participants combined \textit{Lamp A}’s arm with \textit{Lamp B}’s lampshade to form a new lamp structure.
\end{itemize}

The four tasks span a range of interaction granularities, from object-level interaction to component-level modification and recombination. Participants completed the four phases sequentially and proceeded to the next phase only after successfully finishing the current one.

\begin{figure}[t]
  \includegraphics[width=\linewidth]{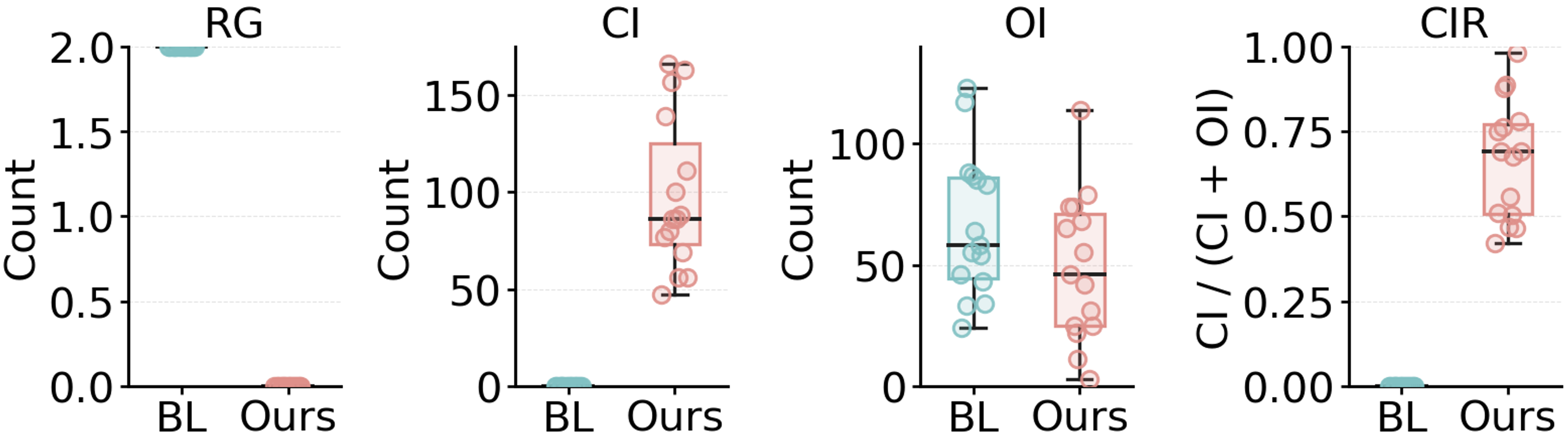}
  \vspace{-5mm}
  \caption{Interaction Log Results on Object Manipulability.}
  \Description{Figure 7 compares interaction logs between the baseline and PartInteractor. PartInteractor reduces regeneration, increases component-level interaction, and produces a higher component interaction ratio than the baseline.}
  \label{fig:study1_result}
\end{figure}

\subsubsection{\textbf{Measures}}

We extracted the following quantitative measures from the interaction logs:

\begin{itemize}[label=$\circ$, leftmargin=2.2em]
    \item \textbf{RG:} Count of whole-object regeneration events.
    \item \textbf{CI:} Count of component-level interaction events.
    \item \textbf{OI:} Count of object-level interaction events.
\end{itemize}

To further quantify interaction granularity, we compute the Component Interaction Ratio 
(\textit{$\textbf{CIR} = \frac{\text{CI}}{\text{CI} + \text{OI}}$}) to capture the extent of fine-grained component-level interaction.

\subsubsection{\textbf{Questionnaires}}

After this study, participants completed a questionnaire to assess perceived control and intent fulfillment:

\begin{itemize}[label=$\circ$, leftmargin=2.2em]
    \item \textbf{Q1}: “\textit{I felt in control when manipulating the generated object.}”
    \item \textbf{Q2}: “\textit{The system allowed me to modify the object in the way I intended.}”
\end{itemize}

\subsubsection{\textbf{Results}}

As shown in Figure~\ref{fig:study1_result}, participants in the baseline condition heavily relied on whole-object regeneration (RG: M=2.0), particularly for structural modification tasks (Tasks 3 \& 4), whereas regeneration was nearly eliminated in PartInteractor (M=0.0). More importantly, we observed a clear shift in interaction patterns. In the baseline condition, all interactions were performed at the object level (OI: M=66.27). In contrast, PartInteractor enabled substantial component-level interaction (CI: M=98.73), accompanied by a reduction in object-level operations (OI: M=48.93). This shift is further reflected in the CIR (M=0.67), indicating that the majority of interactions were performed at the component level.

As shown in Figure~\ref{fig:questionnaire}, participants reported significantly higher perceived control in the part-aware condition (Q1: M=6.20 vs. 3.27; $MD$=2.93; $CI_{95\%}$=[1.96,3.90]; $d_z$=1.68; $p_{\mathrm{Holm}}$=0.000043). Similarly, participants felt better able to modify objects as intended (Q2: M=6.20 vs. 2.80; $MD$=3.40; $CI_{95\%}$=[2.36,4.44]; $d_z$=1.81; $p_{\mathrm{Holm}}$=0.000025). These subjective results are consistent with the log data, where users performed more component-level interactions, suggesting that direct structural manipulation contributes to a stronger sense of control.

These findings support H1 by showing that part-aware representation changes users’ authoring strategies, shifting interaction from whole-object regeneration and object-level manipulation toward direct fine-grained component-level interaction.

\subsection{Study 2: Intent Alignment}

This study investigates \textbf{H2} by examining how different generative workflows affect ambiguity resolution and perceived intent-result alignment.

\subsubsection{\textbf{Experimental Conditions}}

We compare two workflows that differ in when user intent is resolved during the generation process:
(i) \myul{Direct Generation Pipeline} (Baseline): User speech input is parsed by an LLM-based intent module and directly forwarded to the 3D generation backend without intermediate visualization or confirmation.
(ii) \myul{Intent Scaffolding Pipeline} (PartInteractor): User speech input is first processed through an intent scaffolding stage that presents candidate visual references (retrieval- or generation-based) before committing to 3D generation. Users select and confirm a candidate to finalize the request.

The baseline condition reflects existing generative pipelines that immediately commit to 3D generation after interpreting user input. Both conditions used the same generation backend and XR interaction interface to ensure that observed differences arise from the workflow design rather than generation variability.

\begin{figure}[t]
  \includegraphics[width=\linewidth]{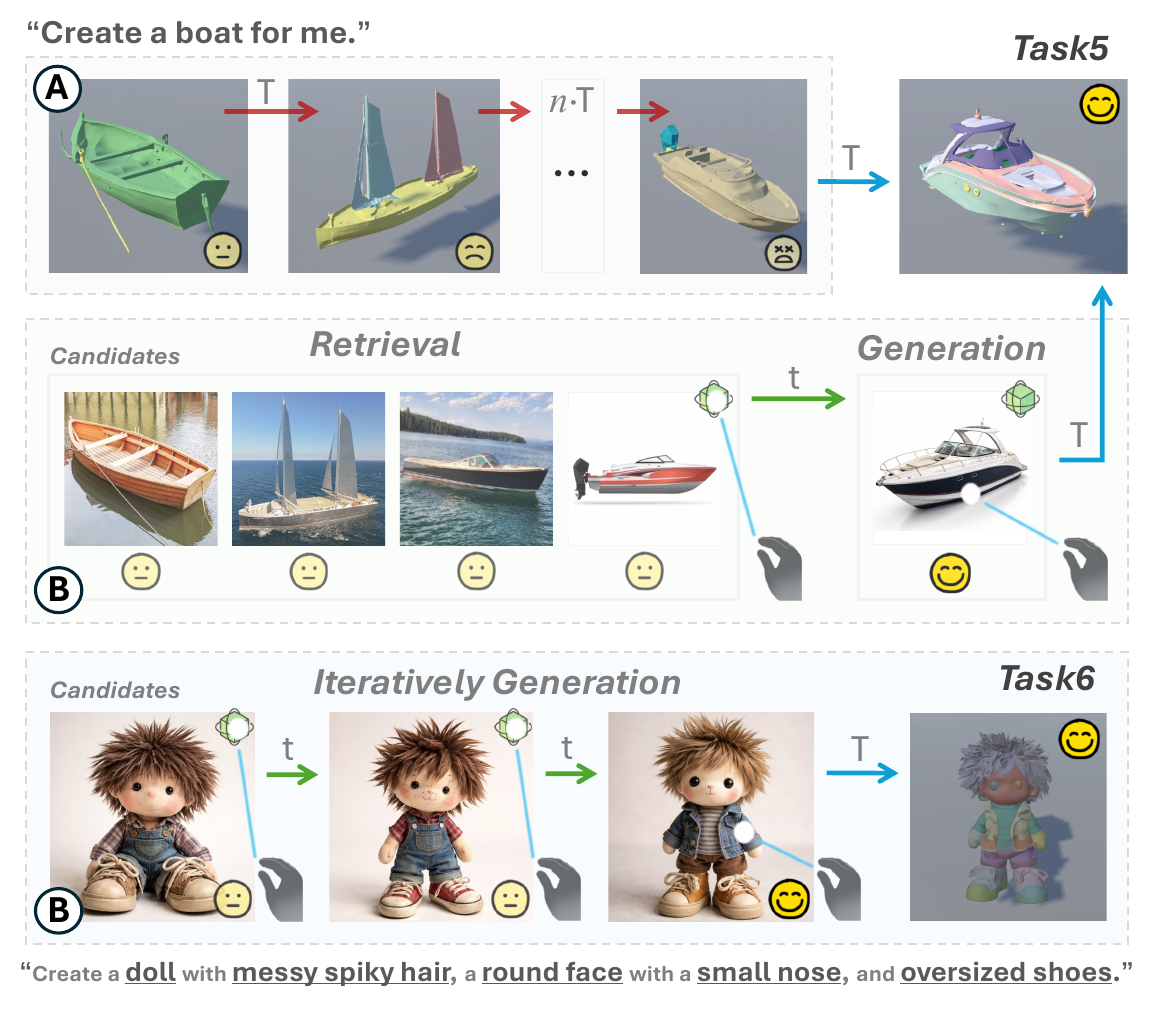}
  \vspace{-6mm}
  \caption{User Study on Intent Alignment. \mycolor{Tasks highlight two object creation workflows: \textcircled{\footnotesize A} Direct Generation baseline, users repeatedly regenerate 3D objects until the result matches their intent. \textcircled{\footnotesize B} Intent Scaffolding pipeline, users first explore retrieved or generated candidate images to clarify their intent before triggering 3D generation. T and t denote the time cost of 3D object and 2D image generation (t<T).}}
  \Description{Figure 8 presents the intent alignment study using two object generation tasks: creating a boat and creating a doll with detailed attributes. It compares two workflows: a direct generation baseline, where users repeatedly regenerate 3D objects until the output matches their intent, and the PartInteractor intent scaffolding pipeline, where users first explore retrieved or generated 2D candidate images before committing to 3D generation.}
  \label{fig:study2}
  \vspace{-2mm}
\end{figure}

\subsubsection{\textbf{Task Design}}

Participants completed two object generation tasks in XR using natural language instructions. These tasks reflect common generative XR workflows, where users describe desired objects and iteratively refine results to match their intent.

\begin{itemize}[leftmargin=12mm]
    \item[\textbf{Task 5}:] \myul{Simple}. “Create a boat for me.”
    \item[\textbf{Task 6}:] \myul{Complex}. “Create a doll with messy spiky hair, a round face with a small nose, and oversized shoes.”
\end{itemize}

The two tasks represent different levels of semantic specificity, allowing us to examine how intent scaffolding supports ambiguity resolution under both underspecified and detailed descriptions. As shown in Figure~\ref{fig:study2}, the two conditions differ in when ambiguity is resolved: either after repeated 3D generation (baseline) or before 3D generation via candidate exploration (PartInteractor). Participants could perform multiple refinements until the generated result matched their intended design, then move to the next task.

\begin{figure}[t]
  \includegraphics[width=0.82\linewidth]{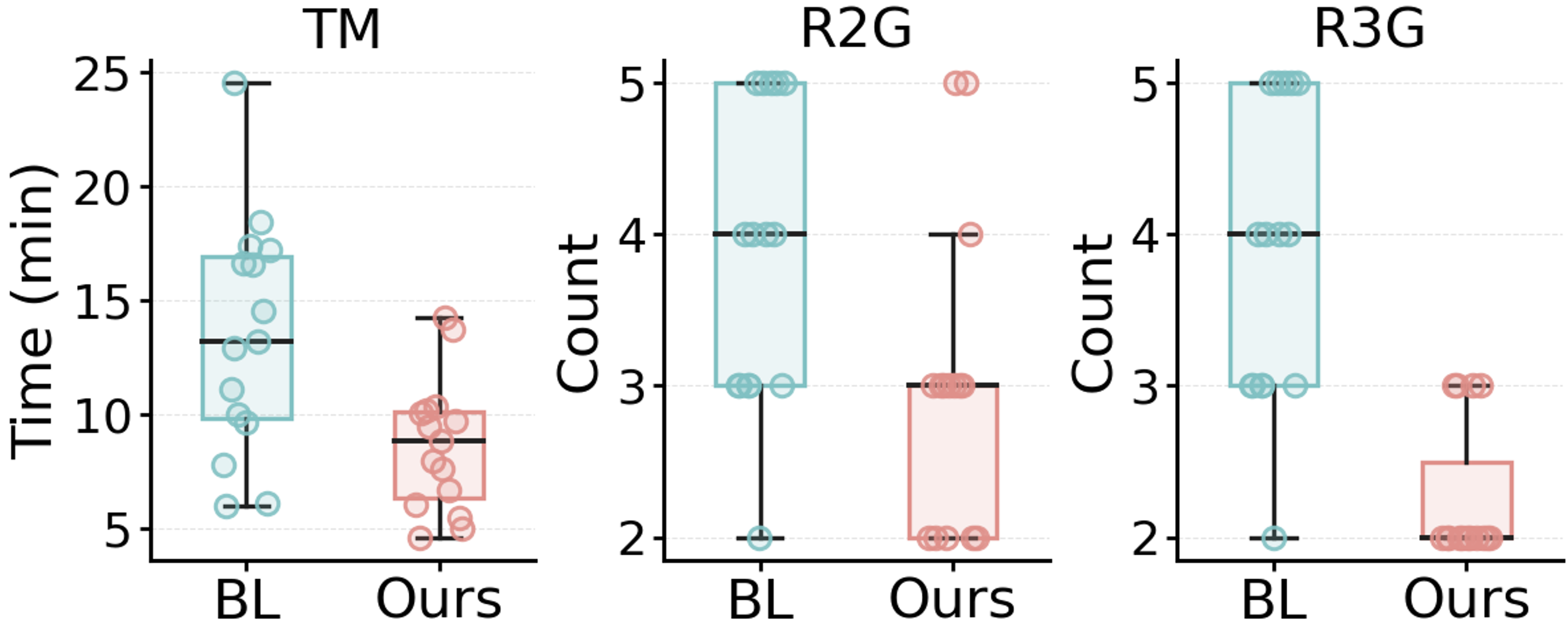}
  \vspace{-1mm}
  \caption{Interaction Log Results on Intent Alignment.}
  \Description{Figure 9 compares intent-alignment logs between the baseline and PartInteractor. PartInteractor reduces task completion time and 3D regeneration attempts while maintaining comparable 2D candidate generation attempts.}
  \label{fig:study2_result}
  \vspace{-2mm}
\end{figure}

\subsubsection{\textbf{Measures}}

We extracted the following quantitative measures from the interaction logs:

\begin{itemize}[label=$\circ$, leftmargin=2.2em]
    \item \textbf{TM:} Time from task start to user-confirmed completion.
    \item \textbf{R2G:} Count of 2D candidate image generation events.
    \item \textbf{R3G:} Count of 3D object generation events.
\end{itemize}


\subsubsection{\textbf{Questionnaires}}

After this study, participants completed a questionnaire to assess intent alignment and interaction quality:

\begin{itemize}[label=$\circ$, leftmargin=2.2em]
    \item \textbf{Q3}: “\textit{The generated result matched what I intended}.”
    \item \textbf{Q4}: “\textit{I rarely needed to regenerate due to system ambiguity}.”
\end{itemize}

\subsubsection{\textbf{Results}}
As shown in Figure~\ref{fig:study2_result}, participants completed tasks faster under the intent scaffolding condition (TM: M=8.68 vs. 13.48). This improvement was accompanied by a reduction in 3D regeneration attempts (R3G: M=2.27 vs. 3.87), indicating that users required fewer iterations to reach satisfactory results. In contrast, the baseline workflow often led to repeated trial-and-error through full 3D regeneration. While both pipelines generate 2D images prior to 3D synthesis, PartInteractor exposes these intermediate results to users, enabling them to preview and refine their intent before committing to 3D generation. Although this involves comparable 2D generation (R2G: M=3.00 vs. 3.87), its lower cost helps reduce unnecessary 3D regeneration and improve overall efficiency.

Subjective results from participants in Figure~\ref{fig:questionnaire} illustrate higher perceived alignment between user intent and generated results (Q3: M=6.13 vs. 3.87; $MD$=2.27; $CI_{95\%}$=[1.32,3.21]; $d_z$=1.33; $p_{\mathrm{Holm}}$=0.000304), along with fewer perceived regeneration needs due to ambiguity (Q4: M=5.80 vs. 3.67; $MD$=2.13; $CI_{95\%}$=[1.09,3.18]; $d_z$=1.13; $p_{\mathrm{Holm}}$=0.000624), suggesting that users were able to resolve ambiguity more effectively and reach satisfactory outcomes with fewer corrective iterations.

These findings support H2 by showing that intent scaffolding helps users resolve ambiguity earlier, shifting refinement from costly repeated 3D regeneration to pre-generation clarification through user-confirmed visual specifications.

\subsection{Study 3: System-Level Co-Creation}

This study investigates \textbf{H3} by examining how participants use PartInteractor in an open-ended XR authoring session.

\begin{figure}[t]
  \includegraphics[width=\linewidth]{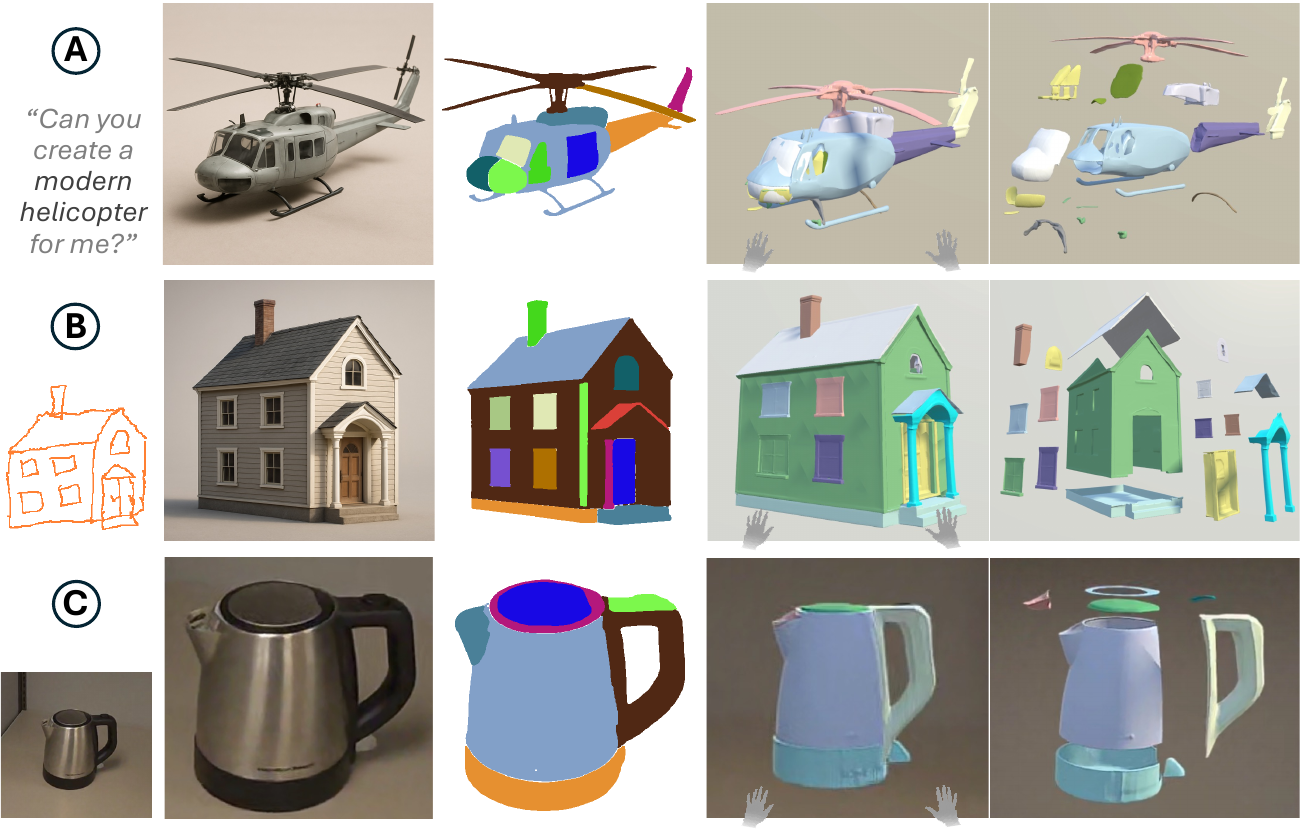}
  \vspace{-6mm}
  \caption{User Study on System-Level Co-Creation. \mycolor{Participants leverage three multimodal authoring inputs: \textcircled{\footnotesize A} speech-based input, where users verbally describe the desired object; \textcircled{\footnotesize B} in-situ sketch input, where users provide a sketch to guide structure; and \textcircled{\footnotesize C} headset-captured image input, where users capture a real-world object as a visual reference for creation.}}
  \Description{Figure 10 presents examples from the system-level co-creation study in an open-ended XR authoring session. Participants use three input modalities to create and refine virtual objects: speech-based input for describing a desired helicopter, in-situ sketch input for guiding the structure of a house, and headset-captured image input for using a real-world kettle as a visual reference. Each row shows the input reference, intermediate generation or part-aware representation, and the resulting 3D object or decomposed components in XR.}
  \label{fig:study3}
  \vspace{-4mm}
\end{figure}

\subsubsection{\textbf{Task Design}}

Participants freely created and edited virtual objects in an open-ended XR authoring session without a predefined goal. Before the session, they were introduced to the available input modalities, including speech, in-situ sketch, and headset-captured image input (Figure~\ref{fig:study3}). Participants could flexibly use any of these modalities, individually or in combination, to express their ideas. They were encouraged to explore, generate objects, and iteratively refine or combine designs they found meaningful. To support exploration, we provided several example categories (e.g., furniture, animals, or objects related to personal interests or daily life). Each session lasted around 15 minutes.

\subsubsection{\textbf{Measures}}

We extracted the following workflow-level measures from the interaction logs:

\begin{itemize}[label=$\circ$, leftmargin=2.2em]
    \item \textbf{ISL} (Interaction Sequence Length): Count of consecutive interactions (CI + OI) applied to an object before the participant switched to generating or selecting a new object, reflecting the depth of within-object refinement.
    \item \textbf{CIR} (Component Interaction Ratio): As defined in Study~1.
\end{itemize}

\subsubsection{\textbf{Questionnaires}}

After this study, participants completed a questionnaire to assess their overall experience:

\begin{itemize}[label=$\circ$, leftmargin=2.2em]
    \item \textbf{Q5}: “\textit{The interaction felt like a continuous design process}.”
    \item \textbf{Q6}: “\textit{The interaction allowed me to stay engaged in the creative process}.”
    \item \textbf{Q7}: “\textit{The available input modalities helped me express my ideas.}”
    \item \textbf{Q8}: “\textit{The system was useful for creating content that felt personally meaningful to me.}”
\end{itemize}

\begin{figure}[t]
  \includegraphics[width=\linewidth]{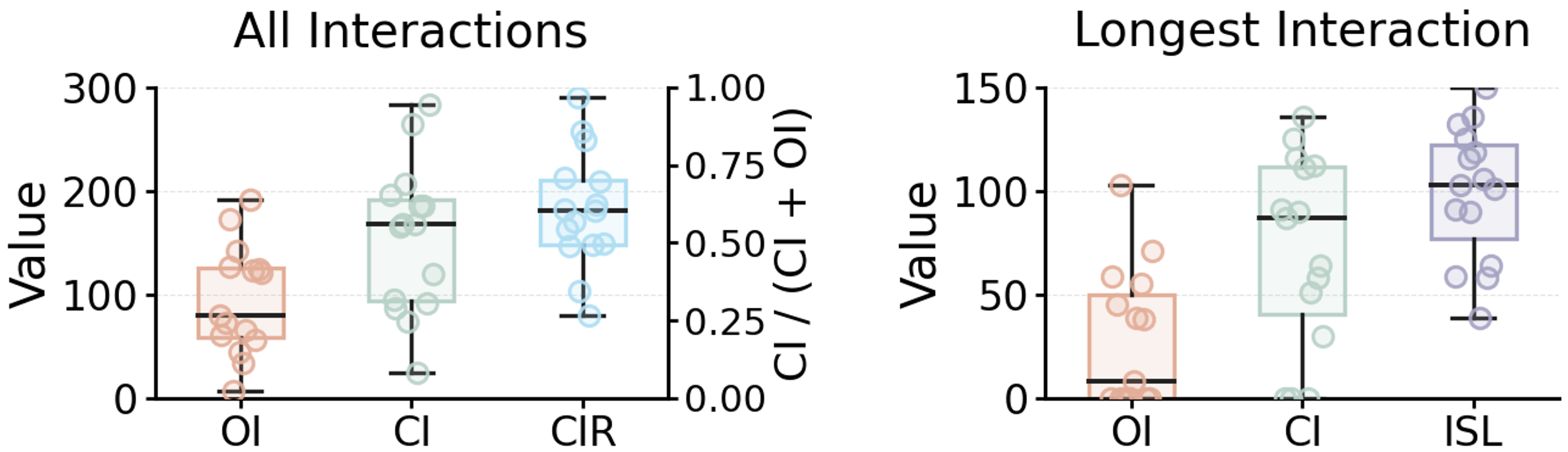}
  \vspace{-6mm}
  \caption{Interaction Log Results on System-Level Co-Creation.}
  \Description{Figure 11 summarizes open-ended co-creation logs. Participants used more component-level than object-level interactions overall and within the longest interaction sequence, showing sustained part-level refinement.}
  \label{fig:study3_result}
  \vspace{-5mm}
\end{figure}

\begin{figure}[t]
  \includegraphics[width=0.86\linewidth]{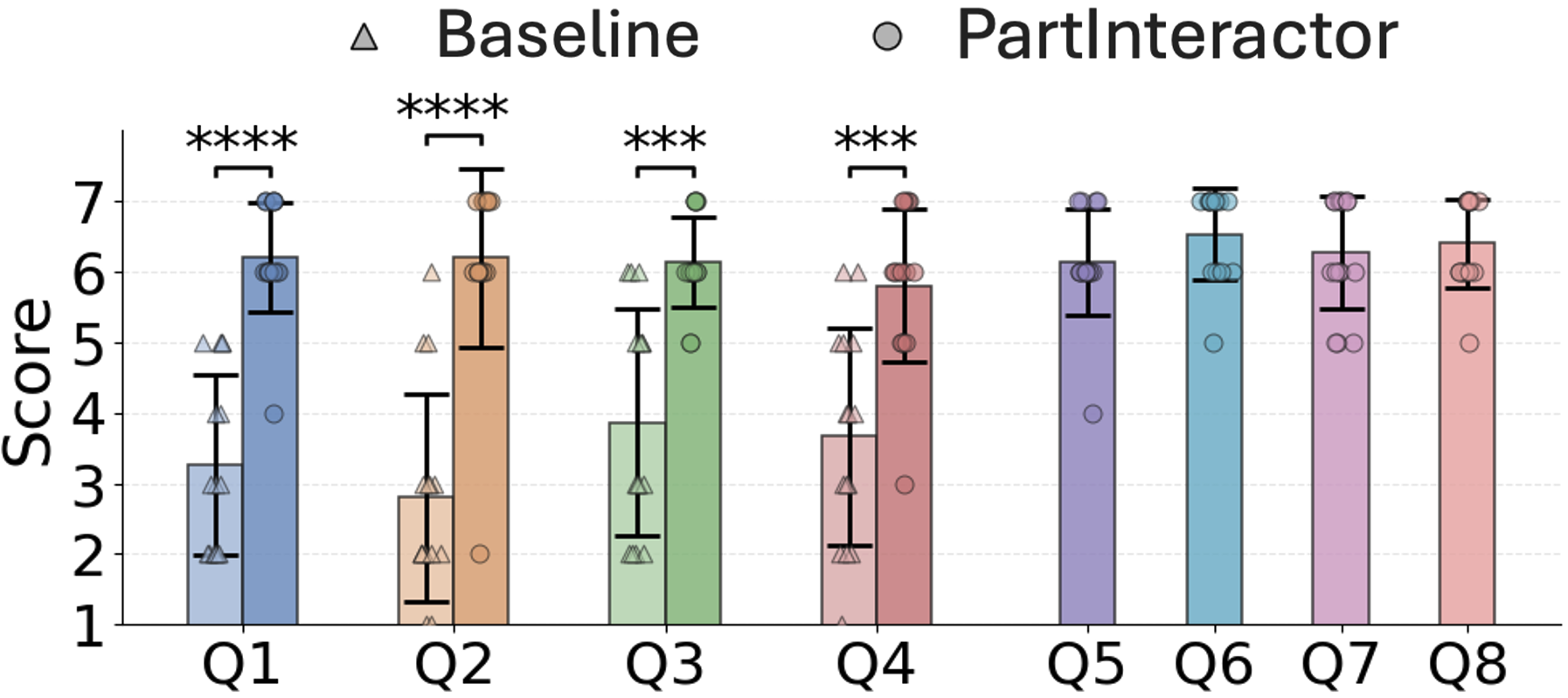}
  \vspace{-2mm}
  \caption{Questionnaire Results (7-point Likert scale) on Object Manipulability (Q1-Q2), Intent Alignment (Q3-Q4), and System-Level Co-Creation (Q5-Q8). Statistical significance: $*$$*$$*$$*$ for $p<0.0001$ and $*$$*$$*$ for $p<0.001$.}
  \Description{Figure 12 summarizes questionnaire results. PartInteractor outperforms the baseline on Q1--Q4, while Q5--Q8 show high ratings for the open-ended co-creation experience.}
  \label{fig:questionnaire}
  \vspace{-3mm}
\end{figure}

\subsubsection{\textbf{Results}}

As shown in Figure~\ref{fig:study3_result}, component-level interactions exceeded object-level interactions across sessions (CI: M=154.47 vs. OI: M=94.87), indicating that users primarily performed fine-grained structural interaction during open-ended authoring. This pattern is further reflected in the Component Interaction Ratio (CIR: M=0.61), showing that most interactions were applied at the component level. We further examined interaction continuity by analyzing the longest interaction sequence on a single object. The Interaction Sequence Length (ISL) reached M=99.27 operations, suggesting sustained refinement within individual objects. Within these sequences, component-level interactions (CI: M=71.40) substantially outnumbered object-level interactions (OI: M=27.87), indicating that long interaction sequences were driven by fine-grained interaction rather than global transformations.

Subjective results in Figure~\ref{fig:questionnaire} further support these observations. Participants reported a strong sense of continuous interaction (Q5: M=6.13) and high engagement during the creative process (Q6: M=6.53). They also indicated that multimodal inputs helped them better express their ideas (Q7: M=6.27), and that the system was effective for creating personally meaningful content (Q8: M=6.40). These consistently high ratings suggest that users perceived the interaction as an ongoing, expressive authoring process.

These findings are consistent with H3, providing exploratory evidence that participants experienced PartInteractor as a continuous and engaging authoring workflow during open-ended use, where they freely created, refined, and recombined generated objects.

\begin{table}[t]
\centering
\caption{Coded themes from post-study interview notes summarizing representative comments and participant counts.}
\Description{Table 2 summarizes five coded themes identified from post-study interviews: overall experience, component-level control, intent externalization, learnability, and system improvements. Participants generally described the experience as positive and engaging, valued the ability to inspect and recombine object components, and found multimodal input helpful for expressing concrete design intent. They also reported a relatively low learning effort. The most frequently mentioned improvement needs concerned generated object quality and speech recognition accuracy, followed by generation speed and contextual understanding. Each representative fragment is accompanied by the number of participants who expressed a similar comment.}
\vspace{-3mm}
\label{tab:interview_themes}
\footnotesize
\setlength{\tabcolsep}{4pt}
\renewcommand{\arraystretch}{1.18}

\begin{tabular}{p{0.26\linewidth}p{0.68\linewidth}}
\toprule
\textbf{Theme} & \textbf{Representative Coded Fragments} \\
\midrule
Overall Experience
& “\textit{good}”~$\times$~2; \, “\textit{great}”~$\times$~6; \, “\textit{interesting}”~$\times$~4; \, “\textit{sense of achievement}”~$\times$~1; \, “\textit{make imagined objects tangible}”~$\times$~5; \, “\textit{entertaining}”~$\times$~1; \, “\textit{engaging}”~$\times$~1. \\
\midrule
Component-Level Control
& “\textit{inspect details}”~$\times$~2; \, “\textit{understand internal structures}”~$\times$~5; \, “\textit{split and inspect component}”~$\times$~2; \, “\textit{combine parts freely}”~$\times$~4; \, “\textit{flexible}”~$\times$~6; \, “\textit{allow creative expression}”~$\times$~2; \, “\textit{operation is smooth}”~$\times$~2. \\
\midrule
Intent Externalization
& “\textit{ambiguity to concrete intent}”~$\times$~3; \, “\textit{drawing is more specific than speaking}”~$\times$~2; \, “\textit{reduce language barrier}”~$\times$~2; \, “\textit{make ideas easier to express}”~$\times$~6; \, “\textit{inspire new designs}”~$\times$~4. \\
\midrule
Learnability
& “\textit{a little, even none}”~$\times$~6; \, “\textit{small effort to learn}”~$\times$~6; \, “\textit{worth to learn}”~$\times$~4; \, “\textit{gain larger than learning cost}”~$\times$~4. \\
\midrule
System Improvements
& “\textit{3D generation speed}”~$\times$~4; \, “\textit{generated objects' quality}”~$\times$~8; \, “\textit{support object copy}”~$\times$~1; \, “\textit{speech recognition accuracy}”~$\times$~8; \, “\textit{system contextual understanding}”~$\times$~2; \, “\textit{context-aware operation suggestions}”~$\times$~1. \\
\bottomrule
\end{tabular}

\vspace{-4mm}
\end{table}

\subsection{Post-Study Interview}

\subsubsection{\textbf{Interview Questions}}
Finally, we conducted a brief semi-structured interview to capture participants’ overall reflections:

\begin{itemize}[label=$\circ$, leftmargin=2.2em]
    \item \textbf{I1}: “\textit{Which workflow did you prefer overall for XR authoring, and why?}”
    \item \textbf{I2}: “\textit{How did component-level interaction affect the way you explored or modified your designs?}”
    \item \textbf{I3}: “\textit{Did the visual or multimodal cues help you clarify or express your intent? In what situations were they most useful?}”
    \item \textbf{I4}: “\textit{Did the system introduce additional learning effort? If so, how did you perceive the trade-off between effort and control?}”
    \item \textbf{I5}: “\textit{Do you have any suggestions for improving the XR authoring interfaces?}”
\end{itemize}

\subsubsection{\textbf{Findings and Implications}}
We analyzed the interview notes using an inductive thematic approach and summarized the coded themes in Table~\ref{tab:interview_themes}. For each theme, we report how many participants mentioned related comments to provide descriptive quantitative coding. Participants' reflections clustered around overall experience, component-level control, intent externalization, learnability, and system improvements.

\textbf{Exposing generated parts supported post-generation control.}
Participants frequently described component-level interaction as useful for inspecting structures, modifying specific parts, and recombining components across objects. This suggests that generated part hierarchies can serve not only as internal generation outputs, but also as editable interaction resources for XR authoring.

\textbf{Visual scaffolding helped users externalize ambiguous intent.}
Participants noted that sketches, generated images, and visual candidates helped turn vague ideas into more concrete authoring targets, especially when language alone was insufficient or affected by recognition errors. This supports intent scaffolding as a useful mechanism to externalize user uncertainty early.

\textbf{Additional interaction effort was acceptable when it increased control.}
Although some participants mentioned additional learning effort, most described it as small or worthwhile when it provided more control, options, or expressiveness. This suggests that extra interaction steps can be acceptable when they clearly reduce uncertainty and improve authoring control.

\textbf{System robustness could enhance practical usability.}
Participants also mentioned limitations such as slow 3D generation, coarse model quality, speech recognition errors, and contextual understanding stability. These comments indicate that the practical system value depends on both interaction design and the reliability of the underlying generative and recognition pipeline.

\begin{figure}[t]
  \includegraphics[width=\linewidth]{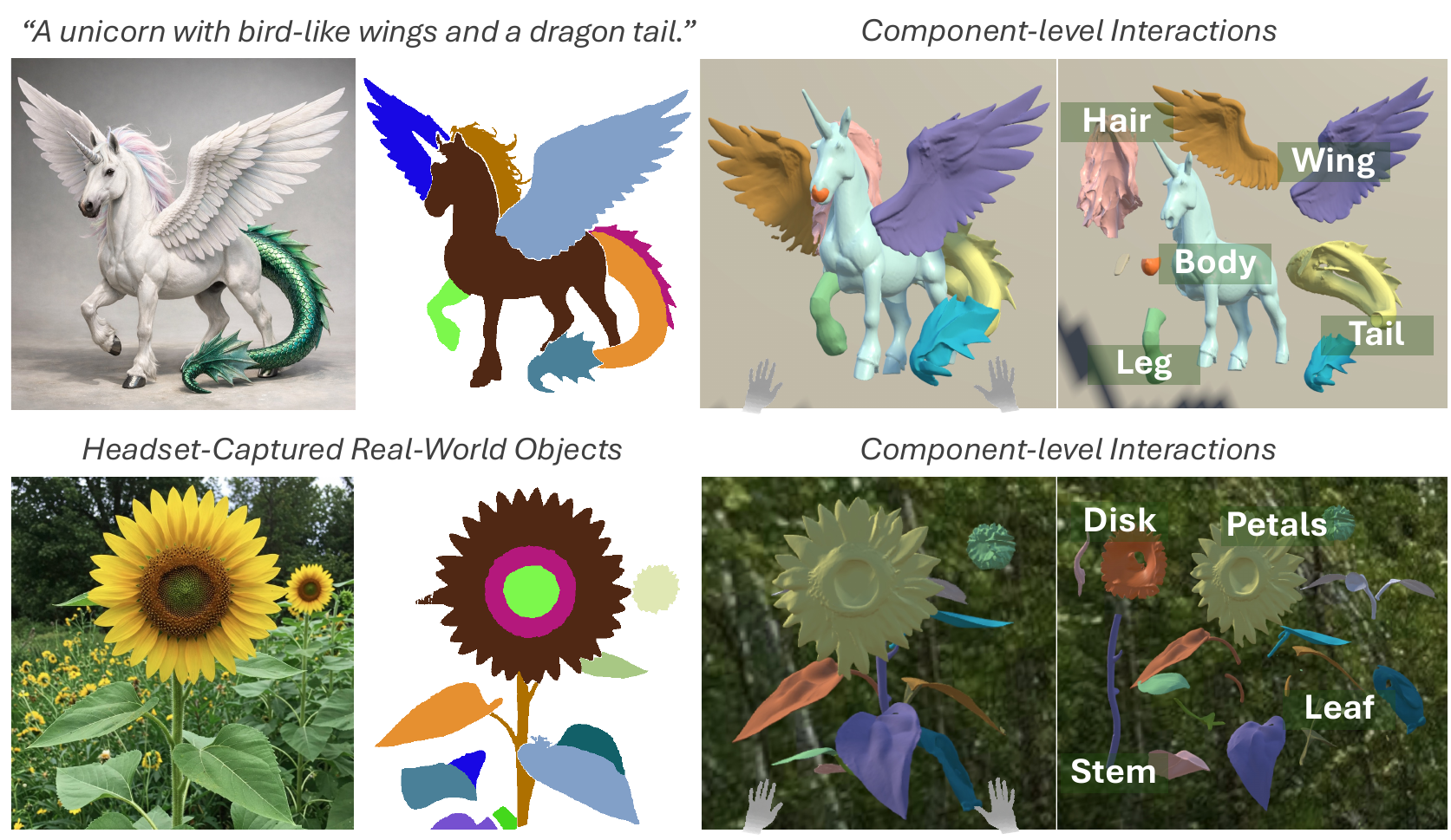}
  \vspace{-6.5mm}
  \caption{Applications in both Virtual and Mixed Reality.}
  \Description{Figure 13 illustrates VR and MR applications of PartInteractor. Users can generate and edit a fantasy unicorn with separable parts, or capture a real-world sunflower and convert it into a component-level virtual object for inspection and manipulation.}
  \label{fig:xr_app}
  \vspace{-3.5mm}
\end{figure}

\section{Applications}

To illustrate the versatility of PartInteractor, we present representative application scenarios across immersive environments, illustrating how intent-driven generation and component-level interaction support a broader range of generative XR authoring workflows.

\vspace{-2mm}
\subsection{Virtual Reality (VR) Environment}

In immersive virtual reality, PartInteractor supports open-ended creative authoring by enabling users to iteratively externalize and manipulate ideas through multimodal interaction. For example, a user designing a virtual artifact or stylized virtual companion may begin with a high-level concept such as “a unicorn with bird-like wings and a dragon tail.” Through intent scaffolding, the system interprets this description into a structured 3D model with identifiable semantic parts. Users can directly manipulate components such as wings, legs, and tail, replacing or recombining them across generated candidates. This transforms object creation from a single-pass generation process into a flexible workflow for iteratively exploring alternative structures through component-level interaction.

Beyond this, such workflows extend to a wide range of creative tasks, including concept prototyping, asset kitbashing, and iterative design exploration, where users benefit from treating generated content as modular and recomposable rather than fixed outputs.

\vspace{-2mm}
\subsection{Mixed Reality (MR) Environment}

In mixed reality, PartInteractor enables situated and interactive authoring by transforming real-world observations into structured, manipulable virtual content. For example, during an outdoor activity, a teacher or student can capture real-world objects in the surrounding environment, such as a flower. The system generates a structured 3D representation that can be placed directly within the shared MR space. Due to preserved semantic components, it can be decomposed and manipulated at the part level to support step-by-step explanation. Petals, stems, and leaves can be separated and examined to illustrate their functional roles. Rather than relying solely on verbal descriptions or static diagrams, users can manipulate the virtual model alongside the real object, helping connect observable appearance with underlying structure and function.

More broadly, this capability supports a variety of scenarios such as interactive teaching, field-based learning, and real-world-inspired content creation, where physical observations can be seamlessly translated into editable and explainable virtual structures.

\section{Discussion}
\subsection{Additional Analysis}

\begin{figure}[t]
  \includegraphics[width=\linewidth]{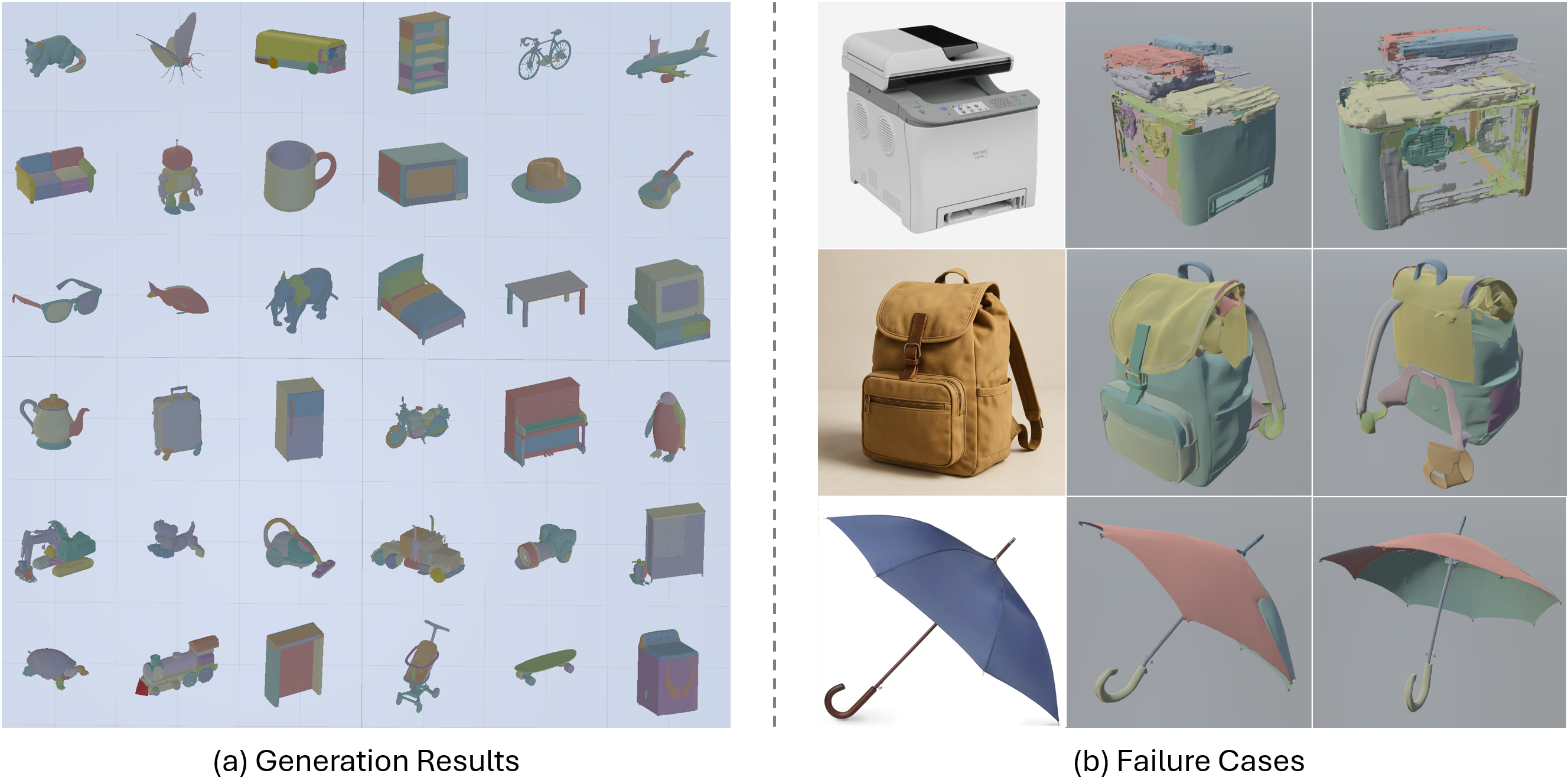}
  \vspace{-6mm}
  \caption{Additional Reliability Analysis and Failures.}
  \Description{Figure 14 shows additional reliability analysis. Most generated objects preserve visually separable part structures, while failure cases include noisy fragments, fused components, and incomplete or incorrectly separated parts.}
  \label{fig:reliability}
  \vspace{-3.5mm}
\end{figure}

\subsubsection{\textbf{Reliability}}
To further examine PartInteractor beyond the objects used in our user studies, we tested 51 everyday categories across three pathways: retrieval-only, generation-only, and retrieval-generation, with 17 cases each. These results provide additional evidence for supporting diverse everyday objects.

We considered a result usable when the generated object preserved major semantic features and exposed key functional parts as separate components for subsequent interaction. Overall, 46/51 cases met this criterion, and Figure~\ref{fig:reliability}~(a) shows representative examples with visually separable and functionally meaningful parts.

The remaining cases revealed several reliability limits, as shown in Figure~\ref{fig:reliability}~(b). Noisy fragments could interfere with part selection, fused components prevented functional parts from being manipulated independently, and occluded regions were sometimes incomplete or merged with nearby geometry.

\subsubsection{\textbf{Runtime Metrics}}
We further logged stage-level runtime across the three pathways. As shown in Table~\ref{tab:runtime}, the main bottleneck is 3D generation, which takes 68.12--75.91s across pathways. 2D generation adds approximately 17.20s when visual candidates are generated, whereas retrieval, intent parsing, and Unity import are comparatively lightweight. The end-to-end latency ranges from 82.65s for retrieval-only to 100.60s for retrieval-generation, reflecting the cost of authoring a part-aware XR object. This cost profile motivates the visual candidate stage, which helps users resolve intent before invoking the slower 3D generation step.

\begin{table}[t]
\centering
\caption{Mean stage-level runtime for three pathways (Retrieval-only | Generation-only | Retrieval-Generation).}
\Description{Table 3 compares the mean stage-level runtime of the retrieval-only, generation-only, and retrieval-generation pathways. Intent parsing, 2D retrieval, and Unity import each require less than one second on average. In contrast, 2D generation takes about 17.20 seconds, while 3D generation is the main runtime bottleneck at about 72.50 seconds on average. The mean end-to-end runtime is 82.65 seconds for retrieval-only, 90.97 seconds for generation-only, and 100.60 seconds for retrieval-generation.}
\vspace{-3mm}
\label{tab:runtime}
\footnotesize
\setlength{\tabcolsep}{3pt}
\renewcommand{\arraystretch}{1.05}
\begin{tabular}{ll c@{\;|\;}c@{\;|\;}c c}
\toprule
\textbf{Stage} & \textbf{Output} & \multicolumn{3}{c}{\textbf{Latency (s)}} & \textbf{Average (s)} \\ \midrule 
Intent Parsing & Structured Command & 0.6764 & 0.7956 & 0.7986 & 0.7569 \\ 
2D Retrieval & Candidate Image & 0.1152 & --- & 0.1137 & 0.1145 \\ 
2D Generation & Candidate Image & --- & 17.4308 & 16.9717 & 17.2013 \\ 
3D Generation & Part-Aware GLB Asset & 73.4696 & 68.1214 & 75.9087 & 72.4999 \\ 
Unity Import & Instantiated XR Object & 0.1089 & 0.0965 & 0.1108 & 0.1054 \\ \midrule 
End-to-end & XR 3D Object & 82.6517 & 90.9681 & 100.6031 & 91.4076 \\
\bottomrule
\end{tabular}
\vspace{-2mm}
\end{table}

\subsection{Limitations and Future Work}
Our work has several limitations that motivate future research. First, PartInteractor depends on the reliability of the underlying part-aware 3D generation model. Although most tested categories yielded usable part hierarchies, failures in geometry and semantic decomposition remain. As a representation-to-interaction framework, PartInteractor could incorporate stronger generation models to improve asset quality and interaction reliability.

Second, the system adopts the default part granularity produced by the generation model, which may not align with users' authoring goals. Future systems could support intent-driven granularity control, generating coarser or finer decompositions as needed.

Third, although generated components are editable in XR, their semantic information is not maintained throughout interaction. Integrating object-component understanding into XR scenes could enable reasoning about part identities and affordances, supporting more expressive and context-aware interactions.

Finally, our evaluation focused on controlled, short-term XR sessions. While this design helped isolate the effects of part-aware representation and intent scaffolding, it may not fully capture long-term or collaborative creative workflows. Future studies could evaluate PartInteractor with broader user groups, longer deployments, and more diverse tasks in ecologically situated settings.

\section{Conclusion}
This work introduced PartInteractor, a multimodal XR authoring system that rethinks generative XR authoring beyond single-shot object synthesis, toward an intent-driven, part-aware workflow for flexible and fine-grained post-generation control. By preserving semantic part hierarchies and integrating an LLM-driven intent-scaffolding pipeline, PartInteractor enables users to clarify intent before generation and directly manipulate generated content afterward at the component level. Across three user studies, our results provide evidence that this workflow improves perceived post-generation control and intent-result alignment. We further illustrate potential VR and MR scenarios, suggesting opportunities for creative design and educational learning. Overall, our findings point to part-aware representation and intent scaffolding as promising design considerations for future generative XR authoring systems. We hope this work encourages further exploration of human-AI co-creation frameworks that balance generative automation with structural transparency, supporting richer and more expressive, controllable XR authoring experiences.


\begin{acks}
This work was supported in part by NSF under grant M3X-2420351.
\end{acks}

\bibliographystyle{ACM-Reference-Format}
\bibliography{main}

@String{Computing = "Computing" }

@String{Computer = "{IEEE} Computer" }

@String{Springer = "Springer-Verlag" }

@inproceedings{sutherland1968head,
  title={A head-mounted three dimensional display},
  author={Sutherland, Ivan E},
  booktitle={Proceedings of the December 9-11, 1968, fall joint computer conference, part I},
  pages={757--764},
  year={1968}
}

@inproceedings{chung1989exploring,
  title={Exploring virtual worlds with head-mounted displays},
  author={Chung, James C and Harris, Mark R and Brooks, Fredrick P and Fuchs, Henry and Kelley, Michael T and Hughes, John and Ouh-Young, Ming and Cheung, Clement and Holloway, Richard L and Pique, Michael},
  booktitle={Three-Dimensional Visualization and Display Technologies},
  volume={1083},
  pages={42--52},
  year={1989},
  organization={SPIE}
}

@misc{apple2023introducing,
  title={Introducing Apple Vision Pro: Apple’s First Spatial Computer},
  author={{Apple Inc.}},
  howpublished={\url{https://www.apple.com/newsroom/2023/06/introducing-apple-vision-pro/}},
  note={Press Release, June 5, 2023},
  year={2023}
}

@misc{meta2025rayban,
  title={Meta Ray-Ban Display: Breakthrough AI Glasses Available Now},
  author={{Meta Inc.}},
  howpublished={\url{https://www.meta.com/blog/meta-ray-ban-display-ai-glasses-connect-2025/}},
  note={Blog, Sep 30, 2025},
  year={2025}
}

@inproceedings{ghamandi2024unlocking,
  title={Unlocking understanding: An investigation of multimodal communication in virtual reality collaboration},
  author={Ghamandi, Ryan Khushan and Kattoju, Ravi Kiran and Hmaiti, Yahya and Maslych, Mykola and Taranta, Eugene Matthew and McMahan, Ryan P and LaViola, Joseph},
  booktitle={Proceedings of the 2024 CHI Conference on Human Factors in Computing Systems},
  pages={1--16},
  year={2024}
}

@inproceedings{numan2025cocreatar,
  title={CoCreatAR: Enhancing authoring of outdoor augmented reality experiences through asymmetric collaboration},
  author={Numan, Nels and Brostow, Gabriel and Park, Suhyun and Julier, Simon and Steed, Anthony and Van Brummelen, Jessica},
  booktitle={Proceedings of the 2025 CHI Conference on Human Factors in Computing Systems},
  pages={1--22},
  year={2025}
}

@inproceedings{han2025spatio,
  title={SpatIO: Spatial Physical Computing Toolkit Based on Extended Reality},
  author={Han, Seung Hyeon and Han, Yeeun and Park, Kyeongho and Lee, Sangjun and Lee, Woohun},
  booktitle={Proceedings of the 2025 CHI Conference on Human Factors in Computing Systems},
  pages={1--22},
  year={2025}
}

@inproceedings{caetano2025graspr,
  title={GraspR: A Computational Model of Spatial User Preferences for Adaptive Grasp UI Design},
  author={Caetano, Arthur and Luo, Yunhao and Sharma, Adwait and Sra, Misha},
  booktitle={Proceedings of the 38th Annual ACM Symposium on User Interface Software and Technology},
  pages={1--16},
  year={2025}
}

@inproceedings{dogan2024augmented,
  title={Augmented object intelligence with xr-objects},
  author={Dogan, Mustafa Doga and Gonzalez, Eric J and Ahuja, Karan and Du, Ruofei and Cola{\c{c}}o, Andrea and Lee, Johnny and Gonzalez-Franco, Mar and Kim, David},
  booktitle={Proceedings of the 37th Annual ACM Symposium on User Interface Software and Technology},
  pages={1--15},
  year={2024}
}

@inproceedings{lee20253d,
  title={3D Sketching+ 2D Generative AI for Car Exterior Design},
  author={Lee, Seung-Jun and Yoon, Jeongche and Lee, Sang-Hyun and Lee, Joon Hyub and Bae, Seok-Hyung},
  booktitle={Proceedings of the 38th Annual ACM Symposium on User Interface Software and Technology},
  pages={1--14},
  year={2025}
}

@article{vaswani2017attention,
  title={Attention is all you need},
  author={Vaswani, Ashish and Shazeer, Noam and Parmar, Niki and Uszkoreit, Jakob and Jones, Llion and Gomez, Aidan N and Kaiser, {\L}ukasz and Polosukhin, Illia},
  journal={Advances in neural information processing systems},
  volume={30},
  year={2017}
}

@article{brown2020language,
  title={Language models are few-shot learners},
  author={Brown, Tom and Mann, Benjamin and Ryder, Nick and Subbiah, Melanie and Kaplan, Jared D and Dhariwal, Prafulla and Neelakantan, Arvind and Shyam, Pranav and Sastry, Girish and Askell, Amanda and others},
  journal={Advances in neural information processing systems},
  volume={33},
  pages={1877--1901},
  year={2020}
}

@article{ho2020denoising,
  title={Denoising diffusion probabilistic models},
  author={Ho, Jonathan and Jain, Ajay and Abbeel, Pieter},
  journal={Advances in neural information processing systems},
  volume={33},
  pages={6840--6851},
  year={2020}
}

@inproceedings{xiang2025structured,
  title={Structured 3d latents for scalable and versatile 3d generation},
  author={Xiang, Jianfeng and Lv, Zelong and Xu, Sicheng and Deng, Yu and Wang, Ruicheng and Zhang, Bowen and Chen, Dong and Tong, Xin and Yang, Jiaolong},
  booktitle={Proceedings of the IEEE/CVF conference on computer vision and pattern recognition},
  pages={21469--21480},
  year={2025}
}

@inproceedings{tang2024lgm,
  title={Lgm: Large multi-view gaussian model for high-resolution 3d content creation},
  author={Tang, Jiaxiang and Chen, Zhaoxi and Chen, Xiaokang and Wang, Tengfei and Zeng, Gang and Liu, Ziwei},
  booktitle={European Conference on Computer Vision},
  pages={1--18},
  year={2024},
  organization={Springer}
}

@inproceedings{xu2024grm,
  title={Grm: Large gaussian reconstruction model for efficient 3d reconstruction and generation},
  author={Xu, Yinghao and Shi, Zifan and Yifan, Wang and Chen, Hansheng and Yang, Ceyuan and Peng, Sida and Shen, Yujun and Wetzstein, Gordon},
  booktitle={European Conference on Computer Vision},
  pages={1--20},
  year={2024},
  organization={Springer}
}

@inproceedings{lee2025pasta,
  title={Pasta: Part-aware sketch-to-3d shape generation with text-aligned prior},
  author={Lee, Seunggwan and Jung, Hwanhee and Koh, Byoungsoo and Huang, Qixing and Yoon, Sang Ho and Kim, Sangpil},
  booktitle={Proceedings of the IEEE/CVF International Conference on Computer Vision},
  pages={18585--18595},
  year={2025}
}

@article{grattafiori2024llama,
  title={The llama 3 herd of models},
  author={Grattafiori, Aaron and Dubey, Abhimanyu and Jauhri, Abhinav and Pandey, Abhinav and Kadian, Abhishek and Al-Dahle, Ahmad and Letman, Aiesha and Mathur, Akhil and Schelten, Alan and Vaughan, Alex and others},
  journal={arXiv preprint arXiv:2407.21783},
  year={2024}
}

@inproceedings{songtang2025unrealllm,
  title={UnrealLLM: Towards Highly Controllable and Interactable 3D Scene Generation by LLM-powered Procedural Content Generation},
  author={SongTang, SongTang and Zhao, Kaiyong and Wang, Lei and Li, Yuliang and Liu, Xuebo and Zou, Junyi and Wang, Qiang and Chu, Xiaowen},
  booktitle={Findings of the Association for Computational Linguistics: ACL 2025},
  pages={19417--19435},
  year={2025}
}

@inproceedings{huang2025fireplace,
  title={Fireplace: Geometric refinements of llm common sense reasoning for 3d object placement},
  author={Huang, Ian and Bao, Yanan and Truong, Karen and Zhou, Howard and Schmid, Cordelia and Guibas, Leonidas and Fathi, Alireza},
  booktitle={Proceedings of the IEEE/CVF conference on computer vision and pattern recognition},
  pages={13466--13476},
  year={2025}
}

@article{mildenhall2021nerf,
  title={Nerf: Representing scenes as neural radiance fields for view synthesis},
  author={Mildenhall, Ben and Srinivasan, Pratul P and Tancik, Matthew and Barron, Jonathan T and Ramamoorthi, Ravi and Ng, Ren},
  journal={Communications of the ACM},
  volume={65},
  number={1},
  pages={99--106},
  year={2021},
  publisher={ACM New York, NY, USA}
}

@article{kerbl20233d,
  title={3D Gaussian Splatting for Real-Time Radiance Field Rendering},
  author={Kerbl, Bernhard and Kopanas, Georgios and Leimk{\"u}hler, Thomas and Drettakis, George},
  journal={ACM Transactions on Graphics},
  volume={42},
  number={4},
  pages={1--14},
  year={2023}
}

@inproceedings{de2024llmr,
  title={Llmr: Real-time prompting of interactive worlds using large language models},
  author={De La Torre, Fernanda and Fang, Cathy Mengying and Huang, Han and Banburski-Fahey, Andrzej and Amores Fernandez, Judith and Lanier, Jaron},
  booktitle={Proceedings of the 2024 CHI Conference on Human Factors in Computing Systems},
  pages={1--22},
  year={2024}
}

@inproceedings{hu2025thing2reality,
  title={Thing2Reality: Enabling Spontaneous Creation of 3D Objects from 2D Content using Generative AI in XR Meetings},
  author={Hu, Erzhen and Li, Mingyi and Hong, Jungtaek and Qian, Xun and Olwal, Alex and Kim, David and Heo, Seongkook and Du, Ruofei},
  booktitle={Proceedings of the 38th Annual ACM Symposium on User Interface Software and Technology},
  pages={1--16},
  year={2025}
}

@inproceedings{zhang2024vrcopilot,
  title={VRCopilot: authoring 3D layouts with generative AI models in VR},
  author={Zhang, Lei and Pan, Jin and Gettig, Jacob and Oney, Steve and Guo, Anhong},
  booktitle={Proceedings of the 37th Annual ACM Symposium on User Interface Software and Technology},
  pages={1--13},
  year={2024}
}

@article{chen2025llmer,
  title={LLMER: Crafting Interactive Extended Reality Worlds with JSON Data Generated by Large Language Models},
  author={Chen, Jiangong and Wu, Xiaoyi and Lan, Tian and Li, Bin},
  journal={IEEE Transactions on Visualization and Computer Graphics},
  year={2025},
  publisher={IEEE}
}

@inproceedings{vachha2025dreamcrafter,
  title={Dreamcrafter: Immersive Editing of 3D Radiance Fields Through Flexible, Generative Inputs and Outputs},
  author={Vachha, Cyrus and Kang, Yixiao and Dive, Zach and Chidambaram, Ashwat and Gupta, Anik and Jun, Eunice and Hartmann, Bj{\"o}rn},
  booktitle={Proceedings of the 2025 CHI Conference on Human Factors in Computing Systems},
  pages={1--13},
  year={2025}
}

@inproceedings{lee2025imaginatear,
  title={Imaginatear: Ai-assisted in-situ authoring in augmented reality},
  author={Lee, Jaewook and Aleotti, Filippo and Mazala, Diego and Garcia-Hernando, Guillermo and Vicente, Sara and Johnston, Oliver James and Kraus-Liang, Isabel and Powierza, Jakub and Shin, Donghoon and Froehlich, Jon E and others},
  booktitle={Proceedings of the 38th Annual ACM Symposium on User Interface Software and Technology},
  pages={1--21},
  year={2025}
}

@inproceedings{numan2024spaceblender,
  title={Spaceblender: Creating context-rich collaborative spaces through generative 3d scene blending},
  author={Numan, Nels and Rajaram, Shwetha and Kumaravel, Balasaravanan Thoravi and Marquardt, Nicolai and Wilson, Andrew D},
  booktitle={Proceedings of the 37th Annual ACM Symposium on User Interface Software and Technology},
  pages={1--25},
  year={2024}
}

@inproceedings{aghel2024people,
  title={How people prompt generative ai to create interactive vr scenes},
  author={Aghel Manesh, Setareh and Zhang, Tianyi and Onishi, Yuki and Hara, Kotaro and Bateman, Scott and Li, Jiannan and Tang, Anthony},
  booktitle={Proceedings of the 2024 ACM Designing Interactive Systems Conference},
  pages={2319--2340},
  year={2024}
}

@inproceedings{jonsson2022cracking,
  title={Cracking the code: Co-coding with AI in creative programming education},
  author={Jonsson, Martin and Tholander, Jakob},
  booktitle={Proceedings of the 14th Conference on Creativity and Cognition},
  pages={5--14},
  year={2022}
}

@inproceedings{zhou2025instructpipe,
  title={Instructpipe: Generating visual blocks pipelines with human instructions and llms},
  author={Zhou, Zhongyi and Jin, Jing and Phadnis, Vrushank and Yuan, Xiuxiu and Jiang, Jun and Qian, Xun and Wright, Kristen and Sherwood, Mark and Mayes, Jason and Zhou, Jingtao and others},
  booktitle={Proceedings of the 2025 CHI Conference on Human Factors in Computing Systems},
  pages={1--22},
  year={2025}
}

@inproceedings{yuan2022wordcraft,
  title={Wordcraft: story writing with large language models},
  author={Yuan, Ann and Coenen, Andy and Reif, Emily and Ippolito, Daphne},
  booktitle={Proceedings of the 27th International Conference on Intelligent User Interfaces},
  pages={841--852},
  year={2022}
}

@inproceedings{kim2023metaphorian,
  title={Metaphorian: Leveraging large language models to support extended metaphor creation for science writing},
  author={Kim, Jeongyeon and Suh, Sangho and Chilton, Lydia B and Xia, Haijun},
  booktitle={Proceedings of the 2023 ACM Designing Interactive Systems Conference},
  pages={115--135},
  year={2023}
}

@article{jiang2024haigen,
  title={HAIGEN: towards human-AI collaboration for facilitating creativity and style generation in fashion design},
  author={Jiang, Jianan and Wu, Di and Deng, Hanhui and Long, Yidan and Tang, Wenyi and Li, Xiang and Liu, Can and Jin, Zhanpeng and Zhang, Wenlei and Qi, Tangquan},
  journal={Proceedings of the ACM on Interactive, Mobile, Wearable and Ubiquitous Technologies},
  volume={8},
  number={3},
  pages={1--27},
  year={2024},
  publisher={ACM New York, NY, USA}
}

@inproceedings{liu2022opal,
  title={Opal: Multimodal image generation for news illustration},
  author={Liu, Vivian and Qiao, Han and Chilton, Lydia},
  booktitle={Proceedings of the 35th Annual ACM Symposium on User Interface Software and Technology},
  pages={1--17},
  year={2022}
}

@inproceedings{wang2024reelframer,
  title={ReelFramer: Human-AI co-creation for news-to-video translation},
  author={Wang, Sitong and Menon, Samia and Long, Tao and Henderson, Keren and Li, Dingzeyu and Crowston, Kevin and Hansen, Mark and Nickerson, Jeffrey V and Chilton, Lydia B},
  booktitle={Proceedings of the 2024 CHI Conference on Human Factors in Computing Systems},
  pages={1--20},
  year={2024}
}

@inproceedings{shen2025ideationweb,
  title={Ideationweb: Tracking the evolution of design ideas in human-ai co-creation},
  author={Shen, Hanshu and Shen, Lyukesheng and Wu, Wenqi and Zhang, Kejun},
  booktitle={Proceedings of the 2025 CHI Conference on Human Factors in Computing Systems},
  pages={1--19},
  year={2025}
}

@inproceedings{suh2025storyensemble,
  title={StoryEnsemble: Enabling dynamic exploration \& iteration in the design process with AI and forward-backward propagation},
  author={Suh, Sangho and Lai, Michael and Pu, Kevin and Dow, Steven P and Grossman, Tovi},
  booktitle={Proceedings of the 38th Annual ACM Symposium on User Interface Software and Technology},
  pages={1--36},
  year={2025}
}

@inproceedings{prasad2025exploring,
  title={Exploring multimodal generative ai for education through co-design workshops with students},
  author={Prasad, Prajish and Balse, Rishabh and Balchandani, Dhwani},
  booktitle={Proceedings of the 2025 CHI Conference on Human Factors in Computing Systems},
  pages={1--17},
  year={2025}
}

@article{anderson2025exploring,
  title={Exploring GenAI technologies within collaborative learning},
  author={Anderson, Emma and Lin, Grace C and Farid, Amelia and Fenech, Mic and Hanks, Brandon and Klopfer, Eric and Doherty, Emily and Hirshfield, Leanne and Ko, Mon-Lin Monica and Foltz, Peter and others},
  year={2025},
  publisher={CSCL 2025 Proceedings}
}

@inproceedings{lee2021human,
  title={A human-ai collaborative approach for clinical decision making on rehabilitation assessment},
  author={Lee, Min Hun and Siewiorek, Daniel P and Smailagic, Asim and Bernardino, Alexandre and Berm{\'u}dez i Badia, Sergi Berm{\'u}dez},
  booktitle={Proceedings of the 2021 CHI conference on human factors in computing systems},
  pages={1--14},
  year={2021}
}

@inproceedings{zhang2024rethinking,
  title={Rethinking human-AI collaboration in complex medical decision making: a case study in sepsis diagnosis},
  author={Zhang, Shao and Yu, Jianing and Xu, Xuhai and Yin, Changchang and Lu, Yuxuan and Yao, Bingsheng and Tory, Melanie and Padilla, Lace M and Caterino, Jeffrey and Zhang, Ping and others},
  booktitle={Proceedings of the 2024 CHI Conference on Human Factors in Computing Systems},
  pages={1--18},
  year={2024}
}

@inproceedings{hu2025gesprompt,
  title={GesPrompt: Leveraging Co-Speech Gestures to Augment LLM-Based Interaction in Virtual Reality},
  author={Hu, Xiyun and Ma, Dizhi and He, Fengming and Zhu, Zhengzhe and Hsia, Shao-Kang and Zhu, Chenfei and Liu, Ziyi and Ramani, Karthik},
  booktitle={Proceedings of the 2025 ACM Designing Interactive Systems Conference},
  pages={59--80},
  year={2025}
}

@inproceedings{hou2025echoladder,
  title={EchoLadder: Progressive AI-Assisted Design of Immersive VR Scenes},
  author={Hou, Zhuangze and Tian, Jingze and Li, Nianlong and Ren, Farong and Liu, Can},
  booktitle={Proceedings of the 38th Annual ACM Symposium on User Interface Software and Technology},
  pages={1--22},
  year={2025}
}

@inproceedings{yang2025omnipart,
  title={Omnipart: Part-aware 3d generation with semantic decoupling and structural cohesion},
  author={Yang, Yunhan and Zhou, Yufan and Guo, Yuan-Chen and Zou, Zi-Xin and Huang, Yukun and Liu, Ying-Tian and Xu, Hao and Liang, Ding and Cao, Yan-Pei and Liu, Xihui},
  booktitle={Proceedings of the SIGGRAPH Asia 2025 Conference Papers},
  pages={1--12},
  year={2025}
}

@article{hurst2024gpt4o,
  title={Gpt-4o system card},
  author={Hurst, Aaron and Lerer, Adam and Goucher, Adam P and Perelman, Adam and Ramesh, Aditya and Clark, Aidan and Ostrow, AJ and Welihinda, Akila and Hayes, Alan and Radford, Alec and others},
  journal={arXiv preprint arXiv:2410.21276},
  year={2024}
}

@misc{gpt2025image1,
  title={GPT Image 1},
  author={{OpenAI}},
  year={2025},
  howpublished={\url{https://platform.openai.com/docs/models/gpt-image-1}},
  note={Accessed: 2026-03-06}
}

@inproceedings{haque2023instruct,
  title={Instruct-nerf2nerf: Editing 3d scenes with instructions},
  author={Haque, Ayaan and Tancik, Matthew and Efros, Alexei A and Holynski, Aleksander and Kanazawa, Angjoo},
  booktitle={Proceedings of the IEEE/CVF international conference on computer vision},
  pages={19740--19750},
  year={2023}
}

@article{xu2024instantmesh,
  title={Instantmesh: Efficient 3d mesh generation from a single image with sparse-view large reconstruction models},
  author={Xu, Jiale and Cheng, Weihao and Gao, Yiming and Wang, Xintao and Gao, Shenghua and Shan, Ying},
  journal={arXiv preprint arXiv:2404.07191},
  year={2024}
}

@inproceedings{zhu2025agentar,
  title={agentar: Creating augmented reality applications with tool-augmented llm-based autonomous agents},
  author={Zhu, Chenfei and Hsia, Shao-Kang and Hu, Xiyun and Liu, Ziyi and Shi, Jingyu and Ramani, Karthik},
  booktitle={Proceedings of the 38th Annual ACM Symposium on User Interface Software and Technology},
  pages={1--23},
  year={2025}
}

@inproceedings{rosenberg2024drawtalking,
  title={Drawtalking: Building interactive worlds by sketching and speaking},
  author={Rosenberg, Karl Toby and Kazi, Rubaiat Habib and Wei, Li-Yi and Xia, Haijun and Perlin, Ken},
  booktitle={Proceedings of the 37th Annual ACM Symposium on User Interface Software and Technology},
  pages={1--25},
  year={2024}
}

@inproceedings{shi2025caring,
  title={Caring-ai: Towards authoring context-aware augmented reality instruction through generative artificial intelligence},
  author={Shi, Jingyu and Jain, Rahul and Chi, Seunggeun and Doh, Hyungjun and Chi, Hyung-gun and Quinn, Alexander J and Ramani, Karthik},
  booktitle={Proceedings of the 2025 CHI conference on human factors in computing systems},
  pages={1--23},
  year={2025}
}

@inproceedings{liu2026agenthands,
  title={AgentHands: Generating Interactive Hand Gestures for Spatially Grounded Agent Conversations in XR},
  author={Liu, Ziyi and Li, David and Zhou, Zhongyi and Kim, David and Du, Ruofei and Qian, Xun},
  booktitle={Proceedings of the 2026 CHI Conference on Human Factors in Computing Systems},
  pages={1--24},
  year={2026}
}

@inproceedings{duan2026justshape,
  title={JustShape: Exploring Co-Speech Gestures for Multimodal LLM-Powered 3D Parametric Modeling},
  author={Duan, Runlin and Chen, Yuzhao and Hu, Yichen and Liu, Ziyi and Zhu, Chenfei and Hu, Xiyun and Ma, Dizhi and Wang, Xinyi and Ramani, Karthik},
  booktitle={Proceedings of the 2026 CHI Conference on Human Factors in Computing Systems},
  pages={1--31},
  year={2026}
}

@inproceedings{hu2023thingshare,
  title={ThingShare: Ad-Hoc digital copies of physical objects for sharing things in video meetings},
  author={Hu, Erzhen and Gr{\o}nb{\ae}k, Jens Emil Sloth and Ying, Wen and Du, Ruofei and Heo, Seongkook},
  booktitle={Proceedings of the 2023 CHI Conference on Human Factors in Computing Systems},
  pages={1--22},
  year={2023}
}

@inproceedings{faruqi2023style2fab,
  title={Style2Fab: functionality-aware segmentation for fabricating personalized 3D models with generative AI},
  author={Faruqi, Faraz and Katary, Ahmed and Hasic, Tarik and Abdel-Rahman, Amira and Rahman, Nayeemur and Tejedor, Leandra and Leake, Mackenzie and Hofmann, Megan and Mueller, Stefanie},
  booktitle={Proceedings of the 36th Annual ACM Symposium on User Interface Software and Technology},
  pages={1--13},
  year={2023}
}

@inproceedings{arslan2025tinkerxr,
  title={TinkerXR: In-Situ, Reality-Aware CAD and 3D Printing Interface for Novices},
  author={Arslan, O{\u{g}}uz and Akdo{\u{g}}an, Artun and Dogan, Mustafa Doga},
  booktitle={Proceedings of the ACM Symposium on Computational Fabrication},
  pages={1--19},
  year={2025}
}

@inproceedings{iyer2025xr,
  title={XR-penter: Material-Aware and In Situ Design of Scrap Wood Assemblies},
  author={Iyer, Ramya and Dogan, Mustafa Doga and Larsson, Maria and Igarashi, Takeo},
  booktitle={Proceedings of the 2025 CHI Conference on Human Factors in Computing Systems},
  pages={1--16},
  year={2025}
}

@inproceedings{mo2019partnet,
  title={Partnet: A large-scale benchmark for fine-grained and hierarchical part-level 3d object understanding},
  author={Mo, Kaichun and Zhu, Shilin and Chang, Angel X and Yi, Li and Tripathi, Subarna and Guibas, Leonidas J and Su, Hao},
  booktitle={Proceedings of the IEEE/CVF conference on computer vision and pattern recognition},
  pages={909--918},
  year={2019}
}

@article{mo2019structurenet,
  title={StructureNet: hierarchical graph networks for 3D shape generation},
  author={Mo, Kaichun and Guerrero, Paul and Yi, Li and Su, Hao and Wonka, Peter and Mitra, Niloy J and Guibas, Leonidas J},
  journal={ACM Transactions on Graphics (TOG)},
  volume={38},
  number={6},
  pages={1--19},
  year={2019},
  publisher={ACM New York, NY, USA}
}

@article{jones2020shapeassembly,
  title={Shapeassembly: Learning to generate programs for 3d shape structure synthesis},
  author={Jones, R Kenny and Barton, Theresa and Xu, Xianghao and Wang, Kai and Jiang, Ellen and Guerrero, Paul and Mitra, Niloy J and Ritchie, Daniel},
  journal={ACM Transactions on Graphics (TOG)},
  volume={39},
  number={6},
  pages={1--20},
  year={2020},
  publisher={ACM New York, NY, USA}
}

@inproceedings{yan2026x,
  title={X-Part: High Fidelity And Structure Coherent Shape Decomposition And Completion},
  author={Yan, Xinhao and Xu, Jiachen and Li, Yang and Ma, Changfeng and Yang, Yunhan and Wang, Chunshi and Zhao, Zibo and Lai, Zeqiang and Zhao, Yunfei and Chen, Zhuo and others},
  booktitle={Proceedings of the IEEE/CVF Conference on Computer Vision and Pattern Recognition},
  pages={27062--27071},
  year={2026}
}

@inproceedings{xue2023ulip,
  title={Ulip: Learning a unified representation of language, images, and point clouds for 3d understanding},
  author={Xue, Le and Gao, Mingfei and Xing, Chen and Mart{\'\i}n-Mart{\'\i}n, Roberto and Wu, Jiajun and Xiong, Caiming and Xu, Ran and Niebles, Juan Carlos and Savarese, Silvio},
  booktitle={Proceedings of the IEEE/CVF conference on computer vision and pattern recognition},
  pages={1179--1189},
  year={2023}
}

@article{wang2026partnext,
  title={Partnext: A next-generation dataset for fine-grained and hierarchical 3d part understanding},
  author={Wang, Penghao and He, Yiyang and Lv, Xin and Zhou, Yukai and Xu, Lan and Yu, Jingyi and Gu, Jiayuan},
  journal={Advances in Neural Information Processing Systems},
  volume={38},
  year={2026}
}

@inproceedings{leng2024hypersdfusion,
  title={Hypersdfusion: Bridging hierarchical structures in language and geometry for enhanced 3d text2shape generation},
  author={Leng, Zhiying and Birdal, Tolga and Liang, Xiaohui and Tombari, Federico},
  booktitle={Proceedings of the IEEE/CVF Conference on Computer Vision and Pattern Recognition},
  pages={19691--19700},
  year={2024}
}

@article{schulz2014design,
  title={Design and fabrication by example},
  author={Schulz, Adriana and Shamir, Ariel and Levin, David IW and Sitthi-Amorn, Pitchaya and Matusik, Wojciech},
  journal={ACM Transactions on Graphics (TOG)},
  volume={33},
  number={4},
  pages={1--11},
  year={2014},
  publisher={ACM New York, NY, USA}
}

@article{zhu2026generative,
  title={When Generative AI Meets Extended Reality: Enabling Scalable and Natural Interactions},
  author={Zhu, Mingyu and Chen, Jiangong and Li, Bin},
  journal={IEEE Internet Computing},
  year={2026},
  publisher={IEEE}
}

\end{document}